# Spectral optical monitoring of 3C390.3 in 1995-2007: I. Light curves and flux variation in the continuum and broad lines

A.I. Shapovalova[1], L.Č. Popović[2,3], A.N.Burenkov[1], V.H. Chavushyan[4], D.Ilić[3,5], W. Kollatschny[6], A. Kovačević[3,5], N.G. Bochkarev[7], L. Carrasco[4], J. León-Tavares[8], A. Mercado[9], J.R. Valdes[4], V.V. Vlasuyk[1], and E. de la Fuente[10]

[1]  Special Astrophysical Observatory of the Russian AS, Nizhnij Arkhyz, Karachaevo-Cherkesia 369167, Russia
[2]  Astronomical Observatory, Volgina 7, 11160 Belgrade 74, Serbia
[3]  Isaac Newton Institute of Chile, Yugoslavia Branch
[4]  Instituto Nacional de Astrofísica, Óptica y Electrónica, Apartado Postal 51, CP 72000, Puebla, Pue. México
[5]  Department of Astronomy, Faculty of Mathematics, University of Belgrade, Studentski trg 16, 11000 Belgrade, Serbia
[6]  Institut für Astrophysik, Friedrich-Hund-Platz 1, Göttingen, Germany
[7]  Sternberg Astronomical Institute, Moscow, Russia
[8]  Aalto University Metsähovi Radio Observatory, Metsähovintie 114, FIN-02540, Kylmälä, Finland
[9]  Universidad Politécnica de Baja California, Mexicali B.C., México
[10] Instituto de Astronomia y Meteorologia, Dpto. de Física CUCEI, Universidad de Guadalajara, Av. Vallarta 2602, C.P. 44130, Guadalajara, Jalisco, México

Received / Accepted

**ABSTRACT**

*Context.* We present the results of the long-term (1995-2007) spectral monitoring of the broad-line radio galaxy 3C 390.3, a well known AGN with double-peaked broad emission lines, usually assumed to be emitted from an accretion disk.
*Aims.* To explore dimensions and structure of the BLR, we analyze the light curves of the broad Hα and Hβ line fluxes and the continuum flux. To detect variations in the BLR, we analyze the Hα and Hβ line profiles, as well as the change in the line profiles during the monitoring period.
*Methods.* We attempt first to find a periodicity in the continuum and Hβ light curves, finding that there is a high probability of measuring quasi-periodical oscillations. Using the line shapes and their characteristics (such as e.g., peak separation and their intensity ratio, or FWHM) of broad Hβ and Hα lines, we discuss the structure of the BLR. We also cross-correlate the continuum flux with Hβ and Hα lines to determine the dimensions of the BLR.
*Results.* During the monitoring period, we found that the broad emission component of the Hα and Hβ lines, and the continuum flux varied by a factor of ≈4-5. We also detected different structure in the line profiles of Hα and Hβ. An additional central component appears to be present and superimposed on the disk emission. In the period of high activity (after 2002), Hβ became broader than Hα and red wing of Hβ was higher than that of Hα. We detected time lags of ~95 days between the continuum and Hβ flux, and about 120 days between the continuum and Hα flux.
*Conclusions.* Variations in the line profiles, as well as correlation between the line and continuum flux during the monitoring period, are consistent with a disk origin of the broad lines and the possible contribution of some additional region and/or some kind of perturbation in the disk.

**Key words.** galaxies: active – galaxies: quasar: individual (3C 390.3) – line: profiles

## 1. Introduction

Active galactic nuclei (AGN) often exhibit variability in the broad emission lines. The region where broad lines are formed (hereinafter BLR – broad line region) is close to the central supermassive black hole and may hold basic information about the formation and fueling of AGN.

A long-term spectral monitoring of the nucleus of some AGN has revealed a time lag in the response of the broad emission lines relative to flux changes in the continuum. This lag depends on the size, geometry, and physical conditions of the BLR. Thus, the search for correlations between the nuclear continuum changes and flux variations in the broad emission lines may serve as a tool for mapping the geometrical and dynamical structure of the BLR (see e.g., Peterson 1993, and reference therein).

During the past decade, the study of the BLR in some objects has achieved considerable success, mainly because of the increasing number of coordinated multiwavelength monitoring campaigns through the international "AGN Watch" program (see e.g., Peterson 1999). Most of the objects, included in the AGN Watch, are radio-quiet Sy1 galaxies and only one, 3C 390.3 (z=0.0561), is a well known broad-line radio galaxy. It is a powerful double-lobed FRII radio-galaxy with a relatively strong compact core. The two extended lobes, at a position angle of 144°, each one with a hot spot, are separated by about 223″ (Leahy & Perley 1991). A faint well-collimated thin jet at P.A. = 37°, connecting the core to the northern lobe, was observed by Leahy & Perley (1995). The VLBI observations at 5 GHz show evidence of the superluminal motion (with v/c~4) in this parsec-scale jet (Alef et al. 1988; Alef et al. 1996).

---

In a similar way to about 10% of radio-loud AGN, 3C 390.3 emits very broad, double-peaked emission lines. A number of possible models have been suggested to explain double-peaked line profiles: supermassive binary black holes (Gaskell 1996), outflowing biconical gas streams (Zheng 1996), or emission from an accretion disk (Perez et al. 1988; Rokaki et al. 1992). The strong variability of this object, in both continuum and emission lines is well known (Barr et al. 1980; Barr et al. 1983; Yee & Oke 1981; Netzer 1982; Clavel & Wamsteker 1987; Veilleux & Zheng 1991; Wamsteker et al. 1997; O'Brien et al. 1998; Dietrich et al. 1998; Shapovalova et al. 2001; Sergeev et al. 2002; Tao et al. 2008; Gupta et al. 2009). The object is also a highly variable X-ray source, which has a spectrum exhibiting a broad Fe Kα line (Inda et al. 1994; Eracleous et al. 1996; Wozniak et al. 1998). During a multiwavelength monitoring campaign in 1995, several, large-amplitude, X-ray flares were observed in 3C 390.3; in one of them, the X-ray flux increased by a factor of 3 in a period of 12 days (Leighly et al. 1997). The X-ray flux also varies on scales of several days (Gliozzi et al. 2003; Gupta et al. 2009), exhibiting greater variation for a smaller energy range (Gliozzi et al. 2003).

In an analysis of IUE spectra of 3C 390.3 obtained during 1978–1986, Clavel & Wamsteker (1987) detected a variability time lag of 50 and 60 days for the broad C IV λ1549 and Lyα emission lines, relative to the UV-continuum changes. However, from a reanalysis of the same spectral data, Wamsteker et al. (1997) derived a lag of 116±60 days for C IV and 143±60 days for Lyα. Furthermore, from the IUE monitoring data from December 1994 to March 1996, O'Brien et al. (1998) measured a corresponding lag of 35 days for the C IV λ1549 emission line and 60 days for Lyα. From the optical monitoring of 3C 390.3 during 1994–1995, Dietrich et al. (1998) derived, however, a time lag of about 20 days for the Balmer lines, i.e., their response with respect to changes in the X-ray continuum. Furthermore, a temporal lag between the optical and the UV or X-ray continua was not detected. From the spectrophotometric monitoring of 3C390.3 in 1995–1999, Shapovalova et al. (2001) (Paper I, hereinafter) found that the lag of the Hβ emission line relative to the continuum flux changes was ~100 days. Sergeev et al. (2002) inferred a lag between the continuum and Hβ variations for the centroid CCF of ~89 days. The UV-bump, usually observed in a large number of Seyfert galaxies, is very weak or even absent in 3C 390.3 (Wamsteker et al. 1997).

Studies of the variations in both the continuum and broad emission-line profiles and their correlations can provide information about the BLR physics (see e.g., Shapovalova et al. 2009). The double-peaked broad lines of 3C390.3 are indicative of accretion disk emission, therefore long period spectral variation in 3C390.3 can help us determine the nature of accretion disks. In this paper, we present the results of the spectral (Hα and Hβ) monitoring of 3C 390.3 during the period between 1995 and 2007. This work proceeds the long-term monitoring program for 3C390.3 and our first results were published in 2001 (Paper I). In this paper, we present spectral data between 1995 and 2007, and we analyze the continuum and Hα,β line variations. The paper is organized as follow: in Sect. 2 we describe our observations and data reduction, in Sect. 3 we present the analysis of the observations, in Sect. 4 we discuss our results, and in Sect. 5 we outline our conclusions.

## 2. Observations and data reduction

### 2.1. Spectral observations

Spectra of 3C 390.3 (during 158 nights) were taken with the 6 m and 1 m telescopes of the SAO RAS (Russia, 1995–2007) and with INAOE's 2.1 m telescope of the "Guillermo Haro Observatory" (GHO) at Cananea, Sonora, México (1998–2007). They were acquired using long-slit spectrographs, equipped with CCD detector arrays. The typical wavelength interval covered was from 4000 Å to 7500 Å, the spectral resolution varied between 5 and 15 Å, and the S/N ratio was > 50 in the continuum near Hα and Hβ. Spectrophotometric standard stars were observed every night. The log of spectroscopic observations is given in Table 1 (available electronically only).

The spectrophotometric data reduction was carried out using either the software developed at SAO RAS or the IRAF package for the spectra obtained in México. The image reduction process included bias and flat-field corrections, cosmic ray removal, 2D wavelength linearization, sky spectrum subtraction, addition of the spectra for every night, and relative flux calibration based on standard star observations.

### 2.2. Absolute calibration (scaling) of the spectra

The standard technique of the flux calibration of spectra (i.e., comparison with stars of known spectral energy distribution) is not precise enough for the study of AGN variability, since even under good photometric conditions, the accuracy of spectrophotometry is not superior to 10%. Therefore we used standard stars only for a relative calibration.

For the absolute calibration, the fluxes of the narrow emission lines are adopted for scaling the AGN spectra, because they are known to remain constant on timescales of tens of years (Peterson 1993).

However, for 3C 390.3 there has been discussion about the variability of the narrow lines. We considered this in more detail in Paper I and found that in the case of 3C 390.3 there is no reliable evidence of the [O III] λλ4959+5007 flux variability in timescales of between months and years. Therefore, spectra were scaled by the [O III] λλ4959+5007 integrated line flux taken to be $1.7 \times 10^{-13}$ ergs s$^{-1}$cm$^{-2}$ (Veilleux & Zheng 1991). To confirm this, we compared photometric fluxes in the V filter with the spectral fluxes (close to observed λ =5384 Å) obtained from spectra scaled by [OIII]λλ4959,5007. Spectra and photometric data were obtained using the aperture 2″ × 6″, and 15″, respectively. Both observations were performed in the same, or in nearby nights (separated by <3 days). The V magnitudes were transformed into fluxes F(V) using the equation (Alen 1973; Wamsteker 1981; Dietrich et al. 1998)

$$\log F(V, 5500) = -0.4V - 8.439. \quad (1)$$

Figure 1 (available electronically only) shows a correlation between photometric and spectral fluxes in the monitoring period. The regression line corresponds to the equation:

$$F(V) = (0.59 \pm 0.07) + (1.25 \pm 0.03)F(\text{sp}).$$

As can be seen from Fig. 1, there is a strong linear relationship between F(V) and F(sp) with the correlation coefficient R=0.99. This indicates that the [OIII] lines do not vary (at least on a level of a few percent) on timescales of ~13 years, i.e. during our monitoring period.

The scaling of the blue spectra was performed using the method of Van Groningen & Wanders (1992) modified by Shapovalova et al. (2004)[1]. This method allowed us to obtain a homogeneous set of spectra with the same wavelength calibration and the same [OIII]$\lambda$5007 flux. Red spectra were scaled to the constant flux value of the narrow emission line [OI]$\lambda$6300, by also applying the modified method of Van Groningen & Wanders (1992) (see also Shapovalova et al. 2004). As a reference, we used a red spectrum obtained with the GHAO 2.1 m telescope during a good photometric night, and accurately scaled using the [OIII]$\lambda$5007 line (details of the procedure are given in Shapovalova et al. 2008).

### 2.3. Unification of the spectral data

To investigate the long-term spectral variability of an AGN, it is necessary to gather a consistent set of spectral data. This requires us, in turn, to correct the line and continuum fluxes for aperture effects (Peterson & Collins 1983), a detailed justification of which is given in Peterson et al. (1995), and will not be repeated here.

The NLR in 3C390.3 is more compact than for most Sy1 galaxies. In narrow-band [OIII] images, the object displays compact nuclear emission without any extended structure (Baum et al. 1988). The results of panoramic two-dimensional spectrophotometry of the nuclear region of this object imply that the [OIII]$\lambda$5007 emission originates in a zone smaller than r < 2 arsec (Bochkarev et al. 1997). Furthermore, Osterbrock et al. (1975) obtained a rather low value of the [OIII] F($\lambda$4363)/F($\lambda$5007) line ratio, implying a moderately high electron density in the NLR (several $10^6 cm^{-3}$). Therefore, the NLR in 3C 390.3 can be considered to be a point source, and we did not correct for aperture effects either the ratio of non-stellar continuum to narrow-line flux or the ratio of the broad to narrow line flux, as the light losses in the slit are similar for these components. In addition, our apertures are $2'' \times 6''$ or $2.5'' \times 6''$, which are greater than the size of the NLR. However, the light contribution to the continuum from the host galaxy does depend on the aperture, and it is necessary to introduce the corresponding corrections. To accomplish this goal, we adopted the relation (see Peterson et al. 1995)

$$F_{con} = F(4959 + 5007)\left[\frac{F_{con}}{F(4959 + 5007)}\right]_{obs} - G,\qquad(1)$$

where $F(4959 + 5007)$ is the absolute flux in the [OIII] doublet and the value in brackets is the continuum to [OIII] line observed flux ratio, where $G$ is an aperture-dependent correction factor that accounts for the host galaxy flux. Since most of our spectra were obtained with the 6 m telescope through an aperture of $2'' \times 6''$, this instrumental setup was adopted as standard (i.e. $G = 0$ by definition). The value of $G$ for the spectra obtained at the 2.1 m telescope (aperture of $2.5'' \times 6.0''$) is $G = 0.046$, while for those obtained at the 1 m telescope (aperture of $8'' \times 19.8''$) $G = 0.321 \pm 0.028$.

We note here that there is the influence of seeing on $G$ (Sergeev et al. 2002). We estimated that if the seeing changes from $1''$ to $3''$ (that was typical for our observations), the $G$ would change at the level of ~5% (see Sergeev et al. 2002). For standard apertures $2'' \times 6''$, and another aperture $2.5'' \times 6''$ that is also very close to the standard one, the estimated $G$ is very small ($G \sim 0.046 \times 10^{-15} erg\ cm^{-2}\ s^{-1}\ A^{-1}$ for aperture $2.5'' \times 6''$, or 3% of the minimal flux). Therefore, in our case, the seeing has a small influence on $G$.

[1] See Appendix A of Shapovalova et al. (2004)

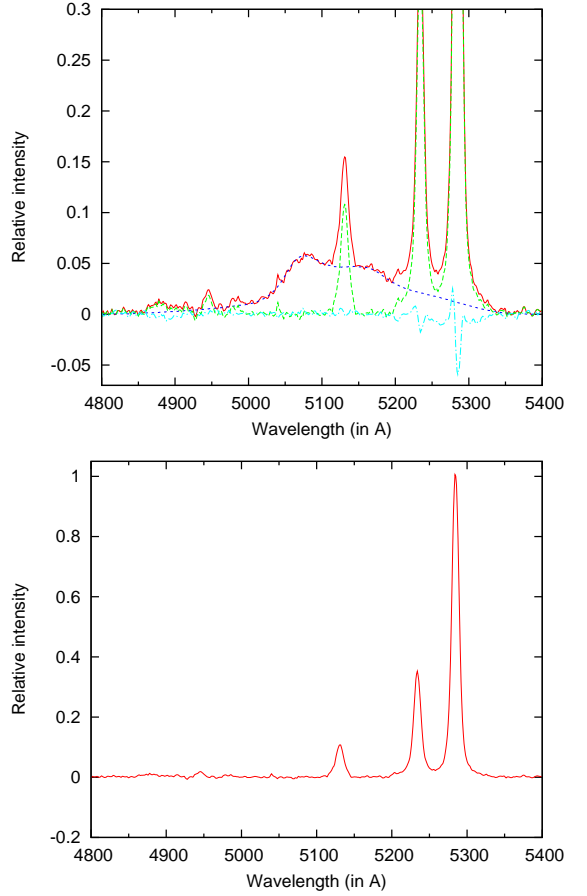

**Fig. 2.** An example of the narrow line template for the H$\beta$ line (bottom) obtained by subtracting from the observed spectrum a spline fit to the broad component (top).

We did not estimate the contribution of the host galaxy, since it is constant and does not influence the analysis of variability, but we mention here that Sergeev et al. (2002) estimated that the host galaxy contributions (for a seeing of $2.5''$) in the H$\alpha$ and H$\beta$ wavelength region are $\sim 7 \times 10^{-16} erg\ cm^{-2} s^{-1} A^{-1}$, and $\sim 4.6 \times 10^{-16} erg\ cm^{-2} s^{-1} A^{-1}$, respectively.

### 2.4. The narrow emission-line template

To obtain only the broad component of Balmer lines, i.e. to remove the narrow lines, we created a narrow line template using the blue and red spectra in the minimum activity state (Sep 09, 1997 and Feb 21, 2002), obtained with a spectral resolution of ~ 8 Å. In these spectra, the broad H$\beta$ component was very weak, and the broad component of the higher Balmer line series were absent. To construct the narrow line template for the H$\beta$ wavelength band, we first estimated the broad component (as can be seen in Fig. 2, top) and then subtracted it from the composite spectrum. The narrow line template is shown in Fig. 2 (bottom).

In the case of H$\alpha$, there is a problem of removing the atmospheric B-band (O2) at $\lambda$=6870 Å in the blue wing. In comparison stars, the B band is very weak due to the shortness of the exposures, therefore it is not possible to remove it using the relative flux calibration. To remove this component, we used the spectra of NGC 4151, taken with long exposures in nights close-by in time. Using different coefficients, we then subtracted the B-band from the H$\alpha$ observed spectrum in the minimum activ-ity state (Sep.9, 1997). Using this spectrum, we fitted the broad component by the spline method (Fig. 3, *panel A*). The narrow line template for the Hα line was obtained by subtracting this broad component from the observed spectrum (Fig. 3, *panel A*).

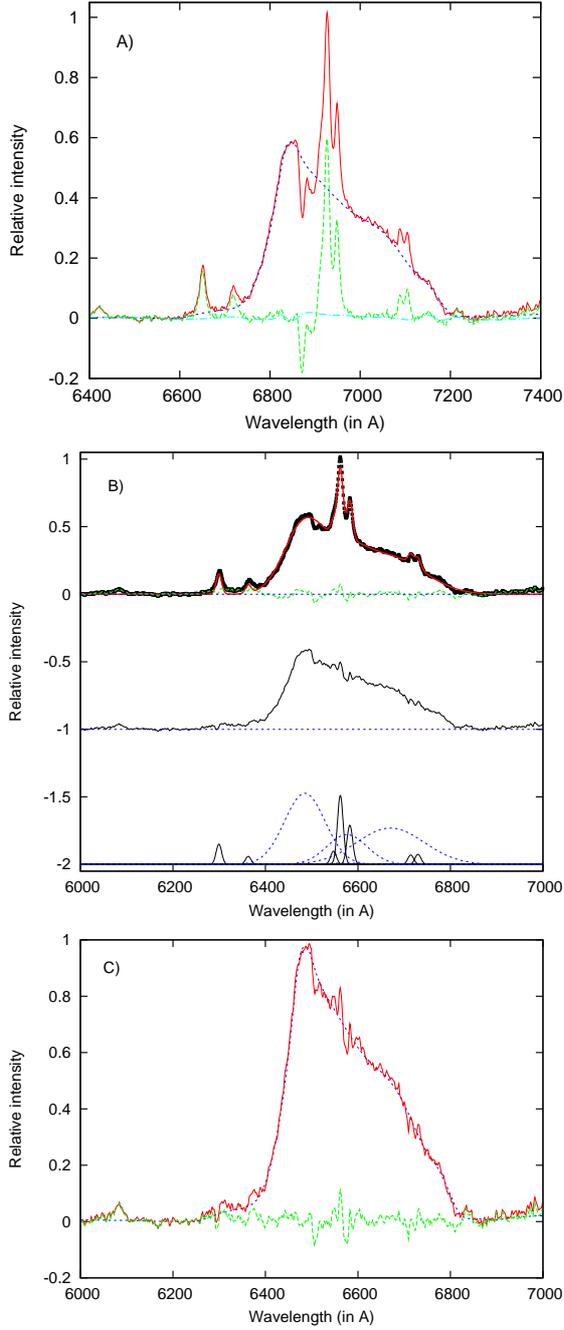

**Fig. 3.** An example of the substraction of the narrow lines in Hα using two different methods: (*panel A*) the observed spectrum (continuous line), the mean broad component (dashed line) and the Hα-narrow-line-template (dotted line); (*panel B*) the observed spectrum and the sum of all Gaussians, as well as their residual (top); the broad component after subtraction of the sum of narrow Gaussians from the observed spectrum (middle); all Gaussian components (bottom); (*panel C*) comparison of the profiles of the broad Hα line obtained by two different methods: the spline (dashed line), the Gaussian-fitting (full line), and the difference between these broad components (dash-dotted line at the bottom).

In addition, we apply the Gaussian-fitting method to the Hα wavelength region, after subtracting the B-band (see Fig. 3, *panel B*). The narrow components are narrow Hα, [N II]λλ 6548, 6584, [O I]λλ 6300, 6364,) [S II]λλ 6717, 6731, while the broad line contains 3 components, the broad blue, the core, and the broad red Gaussians (*panel B*, bottom). In (*panel B*) (up) of Fig. 3, the observed spectrum and the sum of all Gaussian components and their residuals are presented. One can see that the sum of Gaussian components closely represents the Hα observed profile, i.e. the residual is very small. In Fig. 3 (*panel B*, middle), we present the estimated broad component of Hα, obtained by subtraction of sum of narrow Gaussian components, from the observed spectrum (after removing B-band).

We also compare the broad components, obtained by using the spline method and the Gaussian-fitting method (Fig 3, *panel C*). As can be seen from Fig 3 (*panel C*), there is practically no difference between the broad component obtained with these two different procedures.

### 2.5. Measurements of the spectra and errors

From the scaled spectra, we determined the average flux in the continuum at the observed wavelength ∼ 5384Å (or at ∼ 5100Å in the rest frame of 3C390.3, z=0.0556) by means of flux averages in the frame of 5369-5399 Å. To determine the observed fluxes of the broad Hβ and Hα lines, it is necessary to subtract the continuum. To achieve this goal, after subtracting the narrow components, a linear continuum was fitted at windows of 30 Å located at 4800 Å and 5400 Å for the Hβ wavelength band, and at 6560 Å and 7250 Å for the Hα wavelength band.

After the continuum subtraction, we defined the observed fluxes in the lines in the wavelength intervals 4909–5353 Å for Hβ and 6740–7160 Å for Hα.

In Table 2 (available electronically only), the fluxes for the continuum, broad Hα, and Hβ lines are given. The mean error (uncertainty) in the fluxes is for the continuum ≈ 3%, for the broad Hβ line ≈ 5% and the broad Hα line ≈10% (see Table 3, available electronically only). These quantities were estimated by comparing results from spectra obtained within a time interval shorter than 3 days. In addition, we measured the line-part fluxes. To do that, we divided the Hα and Hβ profiles into three parts: the blue wing (covering the blue peak), the core, and the red wing. The intervals adopted were: (i) for Hβ, a blue interval of -9004:-1988 km s⁻¹ (4977Å-5097Å), a central interval of -1988:+1988 km s⁻¹ (5097Å-5165Å), and a red interval of +1988:+8010 km s⁻¹ (5165Å–5268Å); (ii) for Hα, a blue interval of -9007:-1992 km s⁻¹ (6720Å–6882Å), a central interval of -1992:+1992 km s⁻¹ (6882Å–6974Å), and a red interval of +1992:+9007 km s⁻¹ (6974Å–7136Å).

In addition, we measured only the blue and red part of lines by considering the intervals -9004:0:+8010 km s⁻¹ for Hβ and -9007:0:+9007 km s⁻¹ for Hα.

In Table 3, we present the estimated mean flux errors in the Hα and Hβ lines and their segments. Figure 4 (available electronically only) shows the distribution of the flux error in the Hβ line parts against the corresponding fluxes. There is only a weak regression between the flux error and flux in the broad Hβ line, while we can not detect the regression for the case of the broad Hα line.

## 3. Data analysis

We analyzed variations in the continuum and lines using a total of 129 spectra covering the Hβ wavelength region, and 48 spectra covering the Hα line. During the monitoring period, we

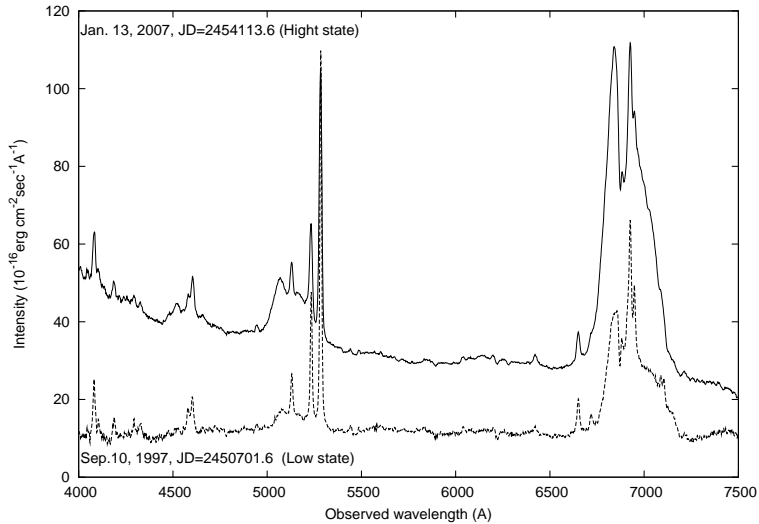

**Fig. 5.** The spectra of 3C390.3 close to maximum (up – observed in 2007) and minimum (bottom – observed in 1997).

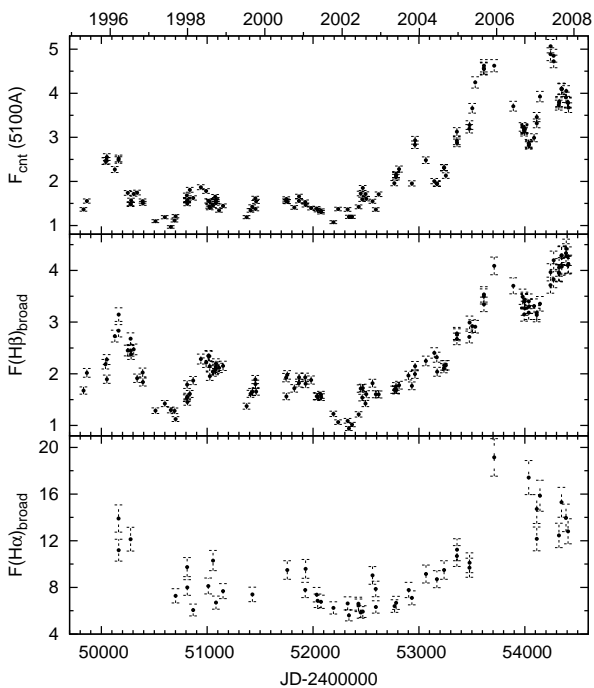

**Fig. 6.** Light curves of the continuum (top), broad Hβ (middle), and Hα (bottom). The continuum flux is given in $10^{-15}\text{erg cm}^{-2}\text{s}^{-1}\text{Å}^{-1}$, and Hβ and Hα line in $10^{-13}\text{erg cm}^{-2}\text{s}^{-1}$.

found that the minimum in the continuum flux occurred during August-September 1997, and the maximum in May-June 2007. In Fig. 5, the shape of 3C390.3 spectra around the minimum and maximum of activity is presented. As can be seen from Fig. 5, the slope of the continuum in the blue part of the spectrum in the minimum-activity state was significantly flatter than in the high-activity state. The wings of Hβ and Hα became extremely weak in the minimum state, and those of Hγ and higher Balmer line series could not be detected at all. These profiles correspond to the Sy 1.8 type and not to Sy 1, as this AGN usually is classified. The spectral type of this object therefore changes with time as noted earlier. In April 1984, the nucleus of 3C390.3 experienced a very deep minimum, the broad wings of hydrogen lines became much weaker (they almost completely vanished in

April 1984) and the spectrum of the nucleus was identified as the Sy 2 (Penston & Perez 1984). Therefore, to compare the continuum and broad line variability we use the broad Hα and Hβ lines only, and the measured continuum around λ5100 Å(rest frame). Our measurements are given in Table 2 (available electronically only).

### 3.1. Variability of the emission lines and the optical continuum

As noted above (see e.g. Paper I, Dietrich et al. 1998; Sergeev et al. 2002; Tao et al. 2008, etc), the optical flux variability of 3C390.3 in the line as well as in the continuum was observed. The light curves of the broad Hα, Hβ line, and continuum fluxes are illustrated in Fig. 6. In general, the light curve of the continuum is similar to that of Hβ and Hα, but there are some differences in the sharpness of the peaks (maxima) and also it is obvious that a lag exists between the continuum and Hβ and Hα light curves.

The maximum amplitude of the flux variations during a monitoring period (1995-2007) corresponds to factors of ∼5.2 for the continuum (at ∼5100Å in the rest frame), ∼4.7 for Hβ, and ∼ 3.4 for Hα (see in §3.5, Table 5). As one can see in Fig. 6, there are several maxima (outbursts). They are more prominent in the continuum light curve than in the Hβ and Hα light curves. In 1995-2002, an outburst of decreasing amplitude and duration (intervals between minima) from 1000 days to ∼400 days can be seen. In 2003-2007, we can discern about ∼3 outbursts on ascending branch of curves of durations ∼600 days, ∼900 days, ∼500 days, and a maximum in 2007. The similar outbursts (oscillations) in the light curve in B band from 1995 to 2001 are also clearly seen (see Fig. 2 in Paper I). These minima/maxima may be indicative of a periodicity (or quasi-periodical oscillation) in the continuum and broad-line light curves.

### 3.2. Light curve analysis

Searching for periodicity has been an important part of variability studies of AGN, because the confirmed periodicity would strongly constrain possible physical models and help us to determine the relevant physical parameters of AGN. We therefore applied a wavelet transform with the Morlet wavelet to see if there is periodicity in the continuum light curve. Wavelet analysis involves a transform from one-dimensional time series to a diffuse two-dimensional time-frequency image for detecting localized (pseudo-)periodic fluctuations from subsets of the time series corresponding to a limited time span (Torrence & Compo 1998).

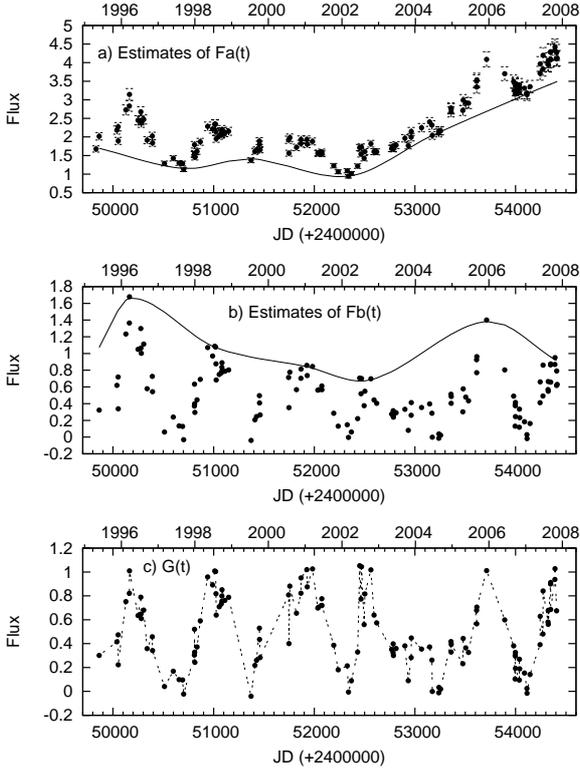

**Fig. 8.** The decomposition of the continuum light curve: a) the flux in the continuum and estimation of $F_A(t)$ component, b) the estimation of $F_B(t)$ component, c) the scaling factor G(t) as an indicator of the quasi-periodical variability. The total flux, $F_A$, and $F_B$ are given in $10^{-15}$erg cm$^{-2}$s$^{-1}$Å$^{-1}$.

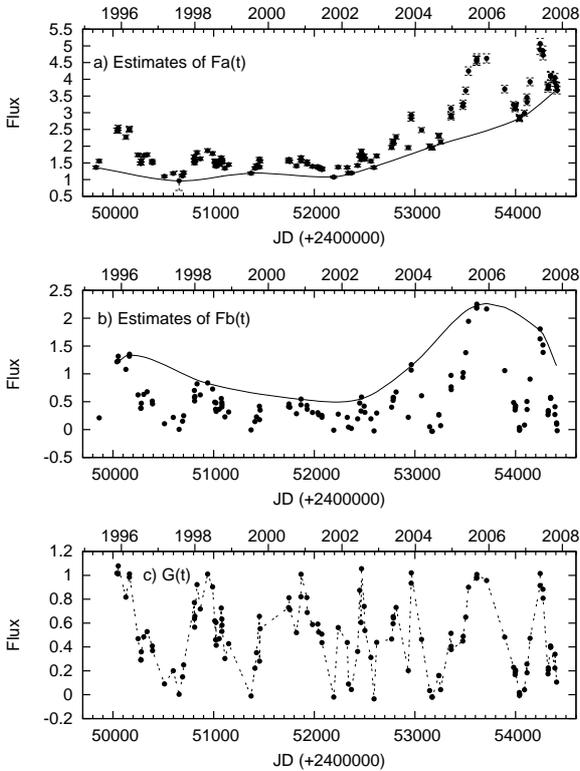

**Fig. 9.** The same as in Fig. 8, but for the H$\beta$ line. The total flux, $F_A$, and $F_B$ are given in $10^{-13}$erg cm$^{-2}$s$^{-1}$.

We employed the standard wavelet IDL codes of Torrence & Compo (1998) to look for periodicities in the 3C 390.3 optical-continuum and H$\beta$ flux light curves. A Morlet wavelet is particularly suited to the analysis of time series and has been successfully applied to study variability in AGN (i.e. Lachowicz et al. 2009; Hovatta 2008; Gupta et al. 2009, etc.). We therefore employ a Morlet wavelet here, which is defined as

$$\Psi_t(s) = \pi^{-1/4} e^{ikt} e^{\frac{-t^2}{2s^2}}, \qquad (2)$$

where $t$ is the time parameter, $\pi^{-1/4}$ is the normalization factor, and $s, k$ are wavelet scale and oscillation frequency parameter, respectively. The parameter $k$ has been set equal to 6 to satisfy the admissibility condition.

The continuous wavelet power spectra is subject to edge arte-facts because the wavelet is not completely localized in time. It is useful to introduce a cone of influence (COI) in which the transform suffers from these edge effects (Torrence & Compo 1998). Periods inside the COI are subject to edge effects and might be dubious.

Figure 7 (available electronically only) shows the continuous wavelet power spectra of the long-term light curve in the optical-continuum and H$\beta$ fluxes, respectively. There are evidently common features in the wavelet power of the two time series. Both wavelet spectra have a long-term periodicity ($\sim 1.65$- 2.45 yr) in the late part of the light curve, above the 95 % confidence level, but within the COI (periodicity being too close to the total signal length).

However, we found a varying timescale of 0.5-0.74 yr around 2004 with a confidence of $> 99.5\%$ that this component does not arise by chance (significance computed assuming the global wavelet spectrum as the theoretical background spectrum following Torrence & Compo (1998)) and located above the COI.

Taking into account that this method provides some indication about periodicity, we performed an analysis assuming that there can be several reasons for variability, e.g. some kinds of shocks or flares in the accretion disk. We therefore assumed that the total flux in the line and continuum can be represented as

$$F_{tot}^{c,l}(t) = F_A^{c,l}(t) + F_B^{c,l}(t) \cdot G^{c,l}(t),$$

where $F_{A,B}^{c,l}$ are different components of the continuum/line fluxes, and $G^{c,l}$ is the scaling function. All these functions may vary with time.

Using the above equation and observed light curves of the continuum and H$\beta$ line, we estimated $F_A(t)$ by assuming this function to be an envelope covering minima on the light curves, and $F_B(t)$ to be an envelope covering maxima on the light curves. The illustration of the decomposition of the continuum and H$\beta$ light curves into three aforementioned functions are given in Figs. 8 and 9.

In Fig. 10 we compare the (rescaled) function $F_A$, $F_B$, and G estimated from the continuum (dotted line, full circles) and H$\beta$ (dashed line, asterisks) light curve. As one can see (Fig. 10, first panel), the shapes of $F_A$ seems to be similar with slight differences, which may be caused by estimation of minima. In contrast, the shapes of $F_B$ are different (second panel), which is indicative of a rapidly increasing continuum $F_B$ from 2002, when higher activity of AGN also started. The $F_B$ component may be quasi-periodical in both the continuum and Hβ, i.e. it may contain two peaks separated by about ten years, that 10-year signal is also visible in the wavelet plot, though in the COI. Finally, the G(t) exhibits (third and fourth panel) quasi-periodical oscillations with a period of around 2-4 yr, which is very close to the period indicated by Morlet wavelet spectral analysis.

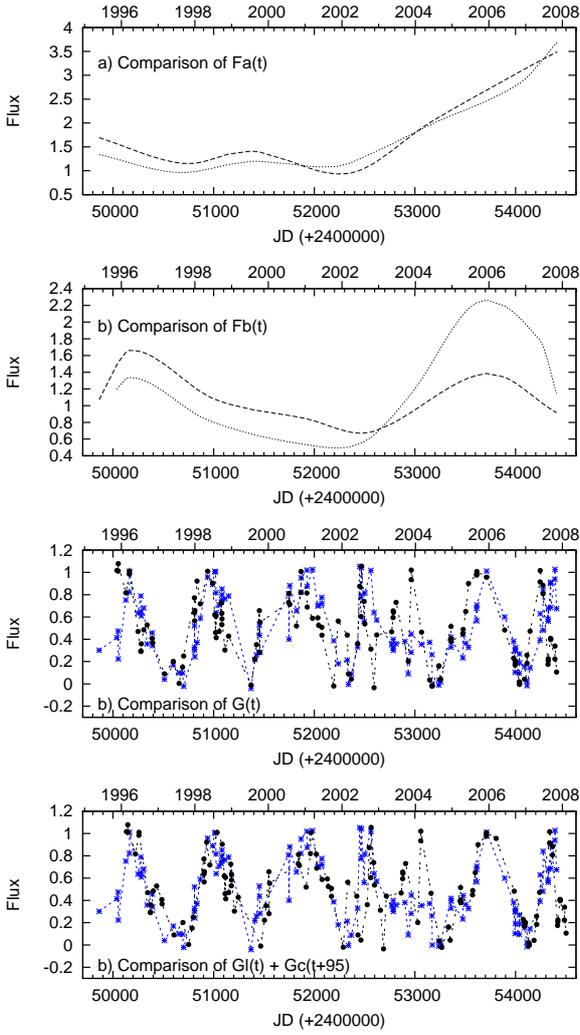

**Fig. 10.** The comparison of different flux components of Hβ (dashed line) and continuum (dotted line). Two last panels present comparison of G(t), where last one is taken to have G(t+95) for the continuum. Asterisks stands for Hβ and full circles for the continuum.

The minima and maxima of the continuum G(t) more closely match those of the Hβ G(t) after a shift of ~95 days, which is close to the lag obtained from the CCF analysis (see Sect. 3.6). It is also interesting that quasi-periodical oscillations can be seen until 2002, before the beginning of the high activity phase (from 2002 to 2007). In the period 1995-2002, there are three prominent maxima. From 2002 to 2004, at the beginning of the activity, a clear peak cannot be seen, while from 2004 to 2007 two maxima can be recognized (see Fig. 10, 3rd and 4th panels).

### 3.3. Changes in the broad line profiles

We first inspected the Hα and Hβ profiles using spectra with a resolution of 8Å for different epochs, comparing Hα and Hβ profiles. As one can see in Fig. 11 (available electronically only), there are no large differences between the line profiles of Hα and Hβ. The lines in all considered epochs show a blue-boosted and red-tailed peaks, which are common for emission from a relativistic disk. However, there are some differences among the Hα and Hβ line profiles that vary during the monitoring period. For instance, Hβ exhibits a bump in its central part during 1997-1998, which later disappears (until 2002). From the maximal phase of activity in 2002, the red wing remains stronger in Hβ than in Hα. During maximum activity, between 2004 and 2008, the Hβ is also broader than Hα.

As noted above, the first inspection of line profiles identified differences between Hα and Hβ line shapes after the beginning of the phase of activity in 2002. We therefore, decided to consider the mean profiles of Hα and Hβ during the whole monitoring period, and both the mean and root-mean-square (rms) profiles for the period before March 05, 2002 (period I, JD2452339.01 according the minimum in Hβ) and after that (period II, see Figs. 12). The mean and rms profiles of both lines have "double-peaked" structure and a full width at zero intensity of about 20 000 km s$^{-1}$. The blue bump in the mean spectrum is located at a radial velocity of about -4000 km s$^{-1}$ relative to the Hα and Hβ narrow components, while the red bump with the flatter peak is located between +2000 and +5000 km s$^{-1}$. The blue bump in the mean profiles is brighter than the red one during 1995-2007. The rms profiles show that the variability of the blue wing is stronger than that of the red (in all periods). As can also be seen in Fig. 12, in the first period a central component was present and apart from the variation in two peaks, there is also a dominant variation in the core of the lines. We note that the sharp narrow central peak in the Hα rms and Hβ rms may be caused by the narrow line subtraction.

We measured the full width at half maximum (FWHM) in the rms and averaged broad line profiles, and we defined an asymmetry parameter A as the ratio of the red to blue half width at half maximum (HWHM), i.e. $A = HWHM_{red}/HWHM_{blue}$. We also estimated the peak positions and their ratios. The measured values for the broad Hβ and Hα lines and their rms values are given in Table 4. As can be seen from Table 4, the FWHM of the mean Hβ line (as well as the rms) obtained from the whole monitoring period is broader than that of Hα, and line (and rms) profiles exhibit the blue asymmetry on the half maximum.

There are also differences between line and rms parameters obtained from periods I and II. As one can see from Table 4, in period I, the FWHMs of Hβ and Hα are practically identical, while Hα exhibits a slightly blue and Hβ slightly red asymmetry (within the errorbars), i.e. they are almost symmetric at the half intensity maximum (here measured to be the blue peak maximum). In contrast, during the phase of flaring (2002-2007), the FWHM of Hβ remains broader than Hα (around 1500 km s$^{-1}$), and both lines have a significant blue asymmetry at half intensity maximum. On the other hand, the separation between peaks is higher in period I than in period II. This result agrees with that of Paper I, where it was found that the difference in the radial velocities of the blue and red peaks is anticorrelated with the brightness of Hβ and continuum. We note that the blue to red peak ratio does not differ too significantly between the two periods, but there is a difference in the peak ratios and their separations of the mean Hα (RP-BP $\approx$ 6500 km s$^{-1}$, Ib/Ir $\approx$ 1.65) and Hβ (RP-BP$\approx$ 5600 km s$^{-1}$, Ib/Ir $\approx$ 1.30) lines, which may reflect a stratification in the BLR, i.e. the different dimensions of the Hα and Hβ emitting regions.

**Table 4.** The line widths and asymmetry measurements for the whole monitoring period, period I and II.

| Spectrum | FWHM km s$^{-1}$ | A | BP km s$^{-1}$ | RP km s$^{-1}$ | Ib/Ir |
|---|---|---|---|---|---|
| H$\alpha_{mean}$ | 10992±499 | 0.85±0.05 | -3709±60 | 2661±240 | 1.65±0.03 |
| H$\alpha_{rms}$ | 9866±723 | 0.63±0.09 | -3807±130 | 1442±80 | 1.52±0.05 |
| H$\beta_{mean}$ | 11918±325 | 0.94±0.00 | -3601±40 | 2154±70 | 1.32±0.02 |
| H$\beta_{rms}$ | 12093±325 | 0.82±0.02 | -3772±40 | 2150±70 | 1.34±0.02 |
| H$\alpha_{mean}$ I | 11526±393 | 0.97±0.03 | -3704±20 | 3178±180 | 1.65±0.01 |
| H$\alpha_{rms}$ I | 10299±149 | 1.08±0.00 | -3632±50 | 4117±30 | 1.02±0.01 |
| H$\beta_{mean}$ I | 11957±504 | 1.05±0.07 | -3453±210 | 2947±770 | 1.41±0.04 |
| H$\beta_{rms}$ I | 12561±610 | 1.10±0.09 | -3408±370 | 4063±20 | 1.12±0.01 |
| H$\alpha_{mean}$ II | 10516±822 | 0.75±0.10 | -3687±10 | 2213±530 | 1.64±0.02 |
| H$\alpha_{rms}$ II | 9362±564 | 0.57±0.06 | -3816±110 | 1450±80 | 1.63±0.03 |
| H$\beta_{mean}$ II | 12015±205 | 0.90±0.01 | -3608±60 | 2151±70 | 1.32±0.02 |
| H$\beta_{rms}$ II | 11373±264 | 0.77±0.00 | -3834±120 | 1671±190 | 1.39±0.01 |

**Notes.** – Col.(1): The averaged and rms broad line profiles of H$\alpha$ and H$\beta$. Col.(2): The measured full width at half maximum (FWHM). Col.(3): The corresponding asymmetry A. Col.(3): The position of the blue peak. Col.(4): The position of the red peak. Col.(5) The ratio of the blue to red peak flux.

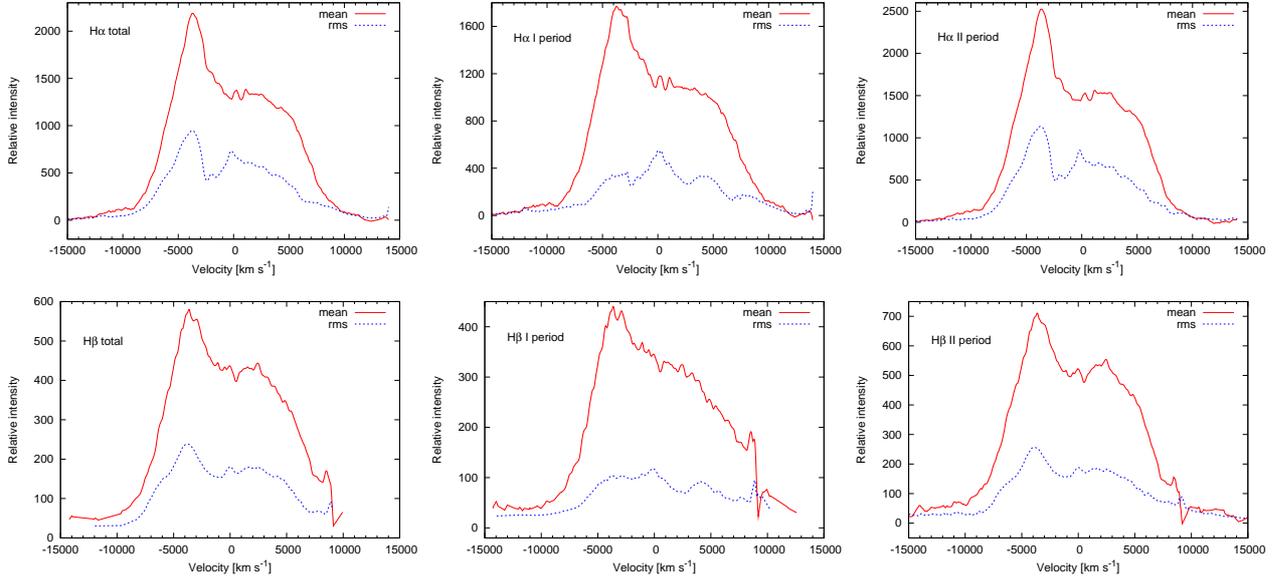

**Fig. 12.** Mean and rms spectra of the H$\alpha$ (top) and H$\beta$ (down) from the whole monitoring period (two left panels) and in the period I (1995-2002, two middle panels) and period II (2002-2007, two right panels).

### 3.4. Flux variability in the wings and core of the broad H$\alpha$ and H$\beta$ lines and various correlations

To determine whether there are any variations in the structure of the BLR (here probably in the disk geometry), we investigated the relationships between different parts of H$\alpha$ and H$\beta$.

Light curves for the wings and cores of H$\alpha$ and H$\beta$ are presented in Fig. 13. As can be seen, the flux in the wings and cores of both lines behaved similarly during the monitoring period. This was also found when we considered only the blue and red parts of the lines.

We found that the flux ratio of different H$\alpha$ to H$\beta$ parts varies during the monitoring period. In Fig. 14, we presented the flux ratios of the red to blue part (asterisk blue-line), the central to blue (circles black-line), and the red to blue for when we divided lines into only two parts (triangle red-line). As one can see from Fig. 14, the blue part remains more prominent in the phase of higher activity in the period 2002-2007. This also reflects some global changes in the line-emitting regions.

We also identified a correlation between the continuum flux and both the integral flux and flux of different parts of the line (see Fig. 15). In Fig. 16 (available electronically only), we present the relationship between the H$\alpha$ and H$\beta$ line parts. As we see in Figs. 15 and 16, there is a linear relationship between the fluxes of the broad H$\alpha$ and H$\beta$ lines, their parts, and continuum fluxes (Fig. 15), as well between the broad line parts themselves (Fig. 16). These linear correlations indicate that for both lines, both the blue and red wings, and the core originate in the BLR that is ionized by the AGN source.

### 3.5. Variability analysis

In Table 5, we present various parameters characterizing the continuum, the integral line, and the line parts, where $F$ is the mean flux and $\sigma(F)$ is its standard deviation, $R$(max) is the ratio of the maximum to the minimum flux in the monitoring period, $F$(var) is a measured (uncertainty-corrected) estimate of the variation

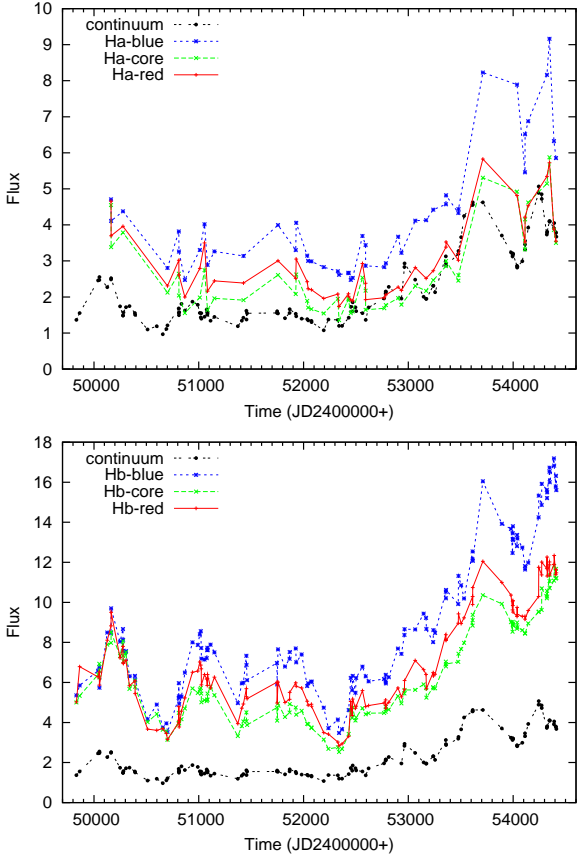

**Fig. 13.** JD- fluxes in the blue and red wings, and core and for Hα (upper panel) and Hβ (bottom panel). The continuum flux is given in $10^{-15}$erg cm$^{-2}$s$^{-1}$Å$^{-1}$, Hβ in $10^{-14}$erg cm$^{-2}$s$^{-1}$ and Hα line in $10^{-13}$erg cm$^{-2}$s$^{-1}$.

amplitude with respect to the mean flux, defined by O'Brien et al (1998) to be

$$F(\text{var}) = [\sqrt{\sigma(F)^2 - e^2}]/F(\text{mean}),$$

where $e^2$ is the mean square value of the individual measurement uncertainty for N observations, i.e. $e^2 = \frac{1}{N}\sum_i^N e(i)^2$. From Table 5, one can see that the amplitude of variability $F(\text{var})$ is larger in the continuum (0.46) than in the Hβ (0.38) and Hα (0.35) broad emission lines. The blue Hβ and Hα wings vary slightly more strongly than the red ones. The maximum to minimum flux ratio, $R(\text{max})$, was maximal in the continuum (5.2) and minimal in Hα (3.4) during the monitoring period. We note that fom the IUE observations performed between 31 December 1994 and 05 March 1996 (O'Brien et al. 1998), $F(\text{var})$ was smaller for Lyα (0.14), CIV (~0.3), and HeII 1640 (0.31) than in optical lines measured during the monitoring period 1995-2007. However, the optical continuum flux variability $F(\text{var})$=0.46 in 1995-2007 was the same as that of the $F(\text{var})$ UV-continuum at wavelength 1370 Å observed in 1994-1996.

### 3.6. Correlations and cross-correlations(CCF) analysis

To estimate the size and structure of the BLR, one can derive the cross-correlation function (CCF) of the continuum light curve with the emission-line light curves. There are several ways to construct a CCF, and it is always advisable to use two or more methods to confirm the obtained results. Therefore, we cross-correlated the 5100 Å continuum light curve with both the Hβ and Hα line (and their parts) light curves using two methods: (i) the z-transformed discrete correlation function (ZDCF) method introduced by Alexander (1997), and (ii) the interpolation cross-correlation function method (ICCF) described by Bischof & Kollatschny (1999).

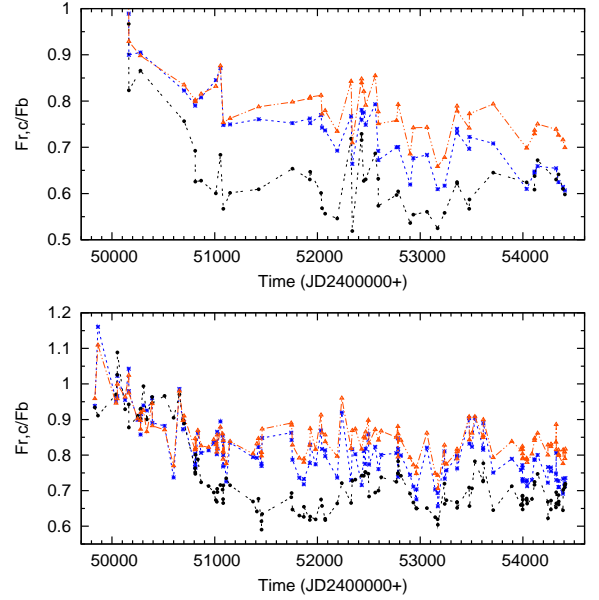

**Fig. 14.** The variation in the ratio red/blue (asterisk blue-line) and central/blue (circles black-line) of the Hα (up) and Hβ (bottom) line component fluxes. The triangle red-line represents the red/blue flux ratio for when we divided lines into only two parts (blue and red).

We choose to apply the CCF method introduced by Alexander (1997), i.e., the z-transformed discrete correlation function (ZDCF). The ZDCF approximates the bin distribution using a bi-normal distribution (see also Shapovalova et al. 2008). In the ZDCF the binning is performed by ensuring that there are equal population, while the effect of measurement errors is estimated by performing the Monte Carlo simulations (usually 100 runs is sufficient) of the averaged ZDCF of light curves with simulated random errors. Since a priori models of light curves do not exist, the estimated ZDCF errors may be extremely large. The Z transforms convergence requires a minimum of 11 points per bin. Using this method, we obtained a lag of $96^{+28}_{-47}$ with coefficient of correlation of 0.92±0.02 between the continuum and Hβ. The cross-correlation functions obtained by ZDCF method are shown in Fig. 17 (available electronically only). The lags and CCF coefficients are given in Table 6. As can be seen for Hα, the CCF is affected by noise because of the small amount of observational data, and it is difficult to reliably determine the lag, there being two peaks in the CCF curve with estimated lags of 24 days $(0.90^{+0.07}_{-0.05})$ and 151 days $(0.90^{+0.06}_{-0.05})$.

As we mentioned above, we applied in addition the ICCF method to cross-correlate the flux of the continuum with both the Hβ and Hα fluxes, and found lags of 97±17 days (CCF=0.95) for Hβ and 270±150 days (CCF=0.84) for Hα, which agree with the ZDCF analysis. Both methods indicate that the lag between the continuum and Hβ flux is around 95 days. For the Hα line, the ICCF method is not applicable (because of the small amount of observations). To confirm the results obtained we compiled centroids for Hα (see Table 6) and measured a lag of 127 ± 18 days, which is closer to the larger value obtained from ZDCF.

**Table 5.** Variability parameters.

| Feature | Region (obs) [Å] | $F$(mean)[a] | $\sigma(F)$[a] | $R$(max)/$R$(min) | $F$(var) |
|---------|------------------|-------------|----------------|-------------------|----------|
| continuum | 5369–5399 | 2.31 | 1.07 | 5.2 | 0.46 |
| H$\beta$ -all | 4909–5353 | 2.39 | 0.91 | 4.7 | 0.38 |
| H$\alpha$-all | 6740–7160 | 9.42 | 3.36 | 3.4 | 0.35 |
| | | | | | |
| H$\beta$-blue wing | 4977–5097 | 0.88 | 0.37 | 4.9 | 0.41 |
| H$\beta$-core | 5097–5165 | 0.63 | 0.24 | 4.7 | 0.38 |
| H$\beta$-red wing | 5165–5268 | 0.70 | 0.26 | 4.3 | 0.37 |
| | | | | | |
| H$\alpha$-blue wing | 6720–6882 | 4.13 | 1.70 | 3.5 | 0.40 |
| H$\alpha$-core | 6882–6974 | 2.63 | 1.16 | 4.3 | 0.42 |
| H$\alpha$-red wing | 6974–7136 | 2.97 | 1.07 | 3.4 | 0.35 |
| | | | | | |
| H$\beta$-blue part | 4977–5131 | 1.20 | 0.48 | 4.8 | 0.39 |
| H$\beta$-red part | 5131–5268 | 1.00 | 0.38 | 4.5 | 0.38 |
| | | | | | |
| H$\alpha$-blue part | 6720–6928 | 5.46 | 2.23 | 3.6 | 0.40 |
| H$\alpha$-red part | 6928–7136 | 4.24 | 1.66 | 3.7 | 0.38 |

**Notes.** – Col.(1): The mean flux. Col.(2): The flux standard deviation. Col.(3): The ratio of the maximum to the minimum flux in the monitoring period. Col.(4): The measured (uncertainty-corrected) estimate of the variation amplitude with respect to the mean flux.
[a] The continuum flux is in units $10^{-15}$erg cm$^{-2}$s$^{-1}$Å$^{-1}$, and the line flux is in $10^{-13}$erg cm$^{-2}$s$^{-1}$.

We also correlated the continuum and H$\beta$ functions $F_A(t)$, $F_B(t)$, and G(t) described in Sect. 3.2 and found no lag between the continuum-H$\beta$ $F_A(t)$, while the lag between the continuum-H$\beta$ $F_B(t)$ is $574^{+12}_{-26}$ days ($0.74^{+0.07}_{-0.06}$). It is interesting that the lag between G$^c$(t) and G$^l$(t) is close to the lag of the entire line within the errorbars and is $97^{+27}_{-46}$ ($0.59^{+0.08}_{-0.07}$). The CCFs between these three function are presented in Fig. 18 (available electronically only).

The calculated lags and CCF between the continuum and different parts of lines, as well as between different parts of H$\alpha$ and H$\beta$ using ZDCF method are given in Table 6. The last two columns of Table 6 indicate the centroid lag and the 90% of the ZDCF maximum, respectively.

In addition, we searched for any linearity between either the H$\beta$ and continuum flux, or the H$\alpha$ and continuum flux, taking different time shifts for the continuum. We assumed that the continuum is shifted by 0, 30, 100, 200, and 300 days. As shown Figs. 19 and 20 (available electronically only), the best-fit linear correlation H$\alpha$ versus (vs.) continuum flux is found when the continuum has a lag of 200 days, but for H$\beta$ the clearest linearity is obtained for a continuum shift of 100 days. This agrees well with the results obtained for the CCF analysis mentioned above.

### 3.6.1. The lines, line-parts, and continuum correlations

We correlated the continuum and line parts, as well as the line parts with the other line parts, finding that the lag in the continuum vs. H$\beta$ line parts is $96^{+47}_{-28}$ in all cases (the CCF coefficient being larger than 0.9). This and a short lag in the H$\beta$ line part - line part lag ($6^{+36}_{-6}$, practically around zero within the errorbars) indicate that line responses to the continuum vary at the same time. In both cases, there is no significant lag between the blue and red H$\beta$ line parts, when we consider only blue and red part, and in the second case when we consider the blue, central, and red part.

For H$\alpha$, Table 6 shows that may be same lag between the blue and red/central part, around 23 days. However, since for

H$\alpha$ there are fewer data available for the calculation than for H$\beta$, as well as that there is one peak around zero (with a slightly smaller CCF value; see values in brackets in Table 6), this should be interpreted with care.

The CCFs for both H$\alpha$ and H$\beta$ in addition to the CCFs between their parts are shown in Fig. 17. On the left side of Fig. 17 (from top to bottom), we present the CCFs of the continuum-total line flux, blue-core, blue-red (three parts), continuum-red, and continuum-blue. On the right side, we present the same for H$\alpha$ line. The CCFs for H$\alpha$ are clearly noisier than for H$\beta$ because there are fewer observations of H$\alpha$ than H$\beta$.

## 4. Discussion of results

We have presented and analyzed data of the long-term monitoring of 3C390.3 (12 years, from 1995 to 2007). In Paper I, we presented the monitoring results for the period 1995–1999, and here we have enlarged our data set for the 8-year period 1999–2007. We now discuss the derived results.

### 4.1. Continuum and line light curves: possible quasi-periodical oscillation

In the first period (1995-2002), three maxima (outbursts) are discernible in the H$\beta$ and continuum light curves, after which in 2002, the activity seems to reach a minimum. Besides these two maxima, there are several local peaks, even in the minimum state, which became enhanced during the maximum of activity (2003-2007). This motivated us to search periodicity in the light curves of both H$\beta$ and continuum (we did not attempt this for H$\alpha$ because there were fewer data than for H$\beta$). By applying the two methods, we showed that there is a high probability that in both the line and continuum light curves some quasi-periodical oscillations exist, i.e., there is some long-term periodicity over between 1.65 and 4 years. Tao et al. (2008) constructed a historical light curve from 1894 to 2004 and identified possible periods of 8.30, 5.37, 3.51, and 2.13 yr.

The appearance of quasi-periodic oscillation (QPOs) is often related to accretion flows around black holes. Chakrabarti et al.

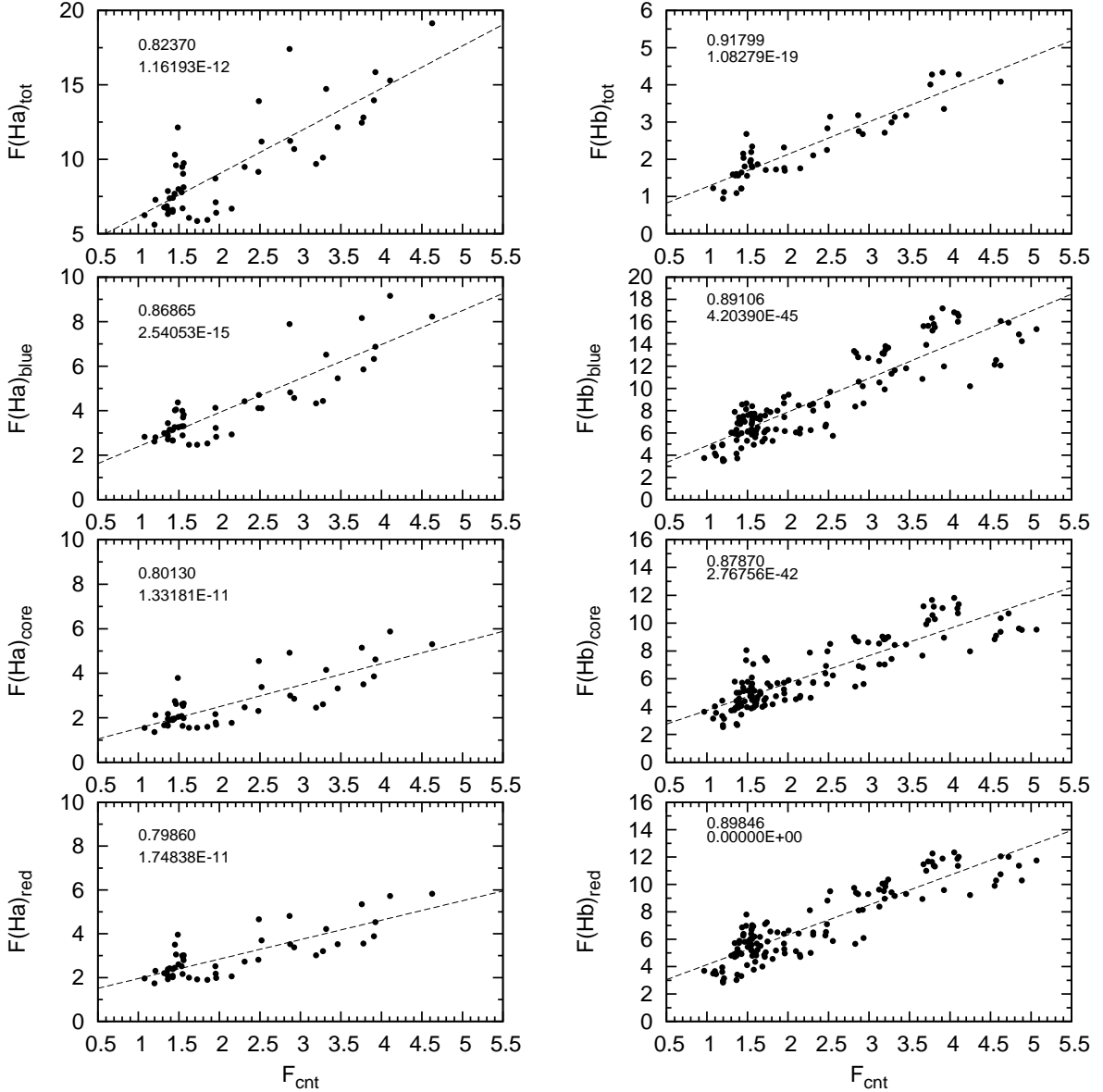

**Fig. 15.** The relationships between the Hα (left) and Hβ (right) total line and line parts with the continuum flux. The continuum flux is given in $10^{-15}$erg cm$^{-2}$s$^{-1}$Å$^{-1}$, Hβ in $10^{-14}$erg cm$^{-2}$s$^{-1}$ and Hα line in $10^{-13}$erg cm$^{-2}$s$^{-1}$. The correlation coefficient and the p-value are given in the upper left corner.

(2004) performed numerical simulations of two-dimensional axisymmetric accretion flows around both stellar and supermassive black holes and found that due to both the radial and vertical oscillation of shock waves in the accretion flow, the quasi-periodic variabilities in the light curve can be expected. These quasi-periodic variabilities may be detectable because of the predominance of the interaction between the winds and the inflow (see Chakrabarti et al. 2004, and references therein). Chakrabarti et al. (2004) found that the QPO frequency of supermassive black hole should be approximately $10^{-6}$–$10^{-7}$ Hz, and the frequencies of possible oscillations were found to be $\sim 10^{-8}$. These variabilities are consistent with a disk geometry. Gierliński et al. (2008) reported that QPO is detected in X-ray spectra of RE J1034+396 (a narrow-line Seyfert 1 galaxy), but the QPO does not appear to have a disk origin (see Middleton et al. 2009).

Harris et al. (2009) also reported the discovery of a X-ray QPO in the M87 jet. Arshakian et al. (2009) found a connec-tion between the variable optical continuum and the subparsecscale jet in 3C390.3, in addition to a significant correlation (at a confidence level of >99.99 %) between the ejected jet components and optical continuum flares and that radio events follow the maxima of optical flares with the mean time delay around ∼36 days. Results obtained in Leon-Tavares et al. (2010) for the broad-line radio galaxy 3C 120, confirm the correlations found in Arshakian et al. (2010) and support the idea that the correlation between optical flares and kinematics of the jet may be common to all radio-loud galaxies and quasars. There is therefore, an open question about the origin of QPOs in 3C390.3.

We note that quasi-periodical variations in the observed blue-to-red line wing flux ratios of Hβ, of period about 10 years, have been reported in several papers (see Veilleux & Zheng 1991; Shapovalova et al. 1996; Bochkarev et al. 1997), maximum values occurring at 1975, 1985, and 1995. From our data (Sect. 3.4, Fig. 14), it follows that the maximal ratio F(blue)/F(red) was ob-

**Table 6.** The time lag and CCF coefficient for the continuum and H$\beta$, $\alpha$ line (total line and parts) using ZDCF method.

| pair I | pair II | lag (ZDCF) (days) | ZDCF | lag (centroid) (days) | 90% Max$_{ZDCF}$ |
|---|---|---|---|---|---|
| $F_{cnt}$ | $F(H\beta)_{tot}$ | $96^{+28}_{-47}$ | $0.92^{+0.02}_{-0.02}$ | $93^{+20}_{-18}$ | 0.83 |
| $F_{cnt}$ | $F(H\beta)_{blue}$ | $96^{+47}_{-28}$ | $0.92^{+0.02}_{-0.02}$ | $71^{+34}_{-26}$ | 0.83 |
| $F_{cnt}$ | $F(H\beta)_{red}$ | $96^{+47}_{-28}$ | $0.92^{+0.02}_{-0.02}$ | $70^{+34}_{-26}$ | 0.83 |
| $F_{cnt}$ | $F(H\beta)_{core}$ | $96^{+47}_{-28}$ | $0.91^{+0.02}_{-0.02}$ | $72^{+33}_{-26}$ | 0.82 |
| $F(H\beta)_{blue}$ | $F(H\beta)_{red}$ | $6^{+36}_{-6}$ | $0.96^{+0.01}_{-0.01}$ | $-5^{+28}_{-30}$ | 0.87 |
| $F(H\beta)_{blue}$ | $F(H\beta)_{core}$ | $6^{+36}_{-6}$ | $0.94^{+0.01}_{-0.01}$ | $-4^{+28}_{-29}$ | 0.85 |
| $F_{cnt}$ | $F(H\alpha)_{tot}$ | $24^{+7}_{-5}$ | $0.90^{+0.05}_{-0.07}$ | $127^{+18}_{-18}$ | 0.81 |
| | | $(151^{+20}_{-9})$ | $0.90(^{+0.05}_{-0.06})$ | | |
| $F_{cnt}$ | $F(H\alpha)_{blue}$ | $64^{+26}_{-21}$ | $0.92^{+0.04}_{-0.06}$ | $76^{+14}_{-13}$ | 0.83 |
| | | $(175^{+25}_{-12})$ | $(0.90^{+0.05}_{-0.06})$ | | |
| $F_{cnt}$ | $F(H\alpha)_{red}$ | $23^{+7}_{-5}$ | $0.87^{+0.07}_{-0.09}$ | $75^{+14}_{-14}$ | 0.79 |
| | | $(175^{+24}_{-12})$ | $(0.81^{+0.08}_{-0.10})$ | | |
| $F_{cnt}$ | $F(H\alpha)_{core}$ | $64^{+26}_{-21}$ | $0.90^{+0.06}_{-0.07}$ | $76^{+14}_{-13}$ | 0.81 |
| | | $(175^{+25}_{-12})$ | $(0.81^{+0.08}_{-0.10})$ | | |
| $F(H\alpha)_{blue}$ | $F(H\alpha)_{red}$ | $23^{+7}_{-5}$ | $0.95^{+0.03}_{-0.04}$ | $10^{+8}_{-9}$ | 0.85 |
| | | $(-1^{+1}_{-8})$ | $(0.93^{+0.02}_{-0.03})$ | | |
| $F(H\alpha)_{blue}$ | $F(H\alpha)_{core}$ | $23^{+7}_{-5}$ | $0.95^{+0.02}_{-0.03}$ | $50^{+13}_{-11}$ | 0.86 |
| | | $(-1^{+1}_{-8})$ | $(0.93^{+0.02}_{-0.03})$ | | |

**Notes.** For a ZDCF performed between the continuum and H$\alpha$, two peaks (within the ZDCF errorbars) are very often present, and the value for the second peak is given in brackets. The last two columns give the centroid lag and the 90% of the ZDCF maximum.

served in 2004–2006. In Figs. 6, 10b, 13, the fluxes of maximal amplitude clearly occur in the years 1996 and 2006, e.g. they divide the interval by about 10 years. This closely agrees with reported quasi-periodical variations of period ~10 years. This period of 10 years may be connected with the BLR (disk) rotation period.

### 4.2. Dimension of the BLR: CCF analysis

In Table 6, the correlations between the continuum flux and either H$\alpha$ or H$\beta$ give different values, indicating different sizes of the H$\alpha$ and H$\beta$ emission regions. The lag measured for H$\beta$ is ~95 days, while for H$\alpha$ it is closer to 100 days. In addition, the coefficient of variability (see Table 5) is slightly higher for H$\beta$ than H$\alpha$, implying that the H$\beta$ emitting region may be more compact than that of H$\alpha$.

In Paper I, the time lag in the response of the emission line relative to flux variations in the continuum, has been found to vary with time i.e. during 1995–1997 a lag of about 100 days was detected, while during 1998–1999, a double-valued lag of $\approx100$ days and $\approx35$ days is present. It is interesting that we obtained a similar result from H$\alpha$, while for H$\beta$ we obtained consistent lag of 95 days using both methods. This, and also difference in the FWHM may indicate multicomponent origin of broad lines, as e.g. in the classical accretion disk, and one disk-like region formed around the accretion disk.

We recall the BLR scenario proposed by Arshakian et al. (2009), where the broad emission lines of a double-peaked structure originate in two kinematically and physically different regions of 3C390.3: 1) BLR1 - a traditional BLR (accretion disk (AD) and the surrounding gas), which is at the distance ~30 light days from the nuclei; 2) BLR2 - a subrelativistic outflow surrounding the jet that is in the cone that is within ~100 light days of the source, at a distance of ∼0.4 pc from the central engine. During the maximal-brightness periods of the nucleus, most of the continuum variable radiation is emitted from the jet and ionizes the surrounding gas, creating the BLR2 than that produces the broad line emission. During the brightness minima, the jet contribution to the ionizing continuum diminishes and the main broad-line emission originates in the "classical" BLR1 (AD), ionized by nuclei continuum related to accretion onto the BH.

One can speculate that this scenario can be used to explain different lags of Hα and Hβ (and difference in the rate of variation), but there may also be some perturbation in a "classical" accretion disk in combination with the different dimensions of the emitting regions that may explain the observed results. In a forthcoming paper, we propose to apply disk models to explore these findings.

### 4.3. The structure of the BLR - disk and perturbations

An accretion disk is usually assumed to produce the double-peaked line profiles emitted by some AGN. Gezari et al. (2007) found, from their approximately 20-year long spectral monitoring of the double-peaked Hα emission line of seven broad-line radio galaxies, that their profiles varied on timescales of from months to years, and double-peaked lines were successfully modeled by emission of the gas from the outer part of the accretion disk. To explain variable flux ratios of the line wings, more complex models are required, such as hot spots (Zheng 1996), two-arm spiral waves in the accretion disk (Chakrabarti & Wiita 1994), or a relativistic eccentric disk (Eracleous et al. 1995). Therefore, accretion-disk emission from 3C390.3 is probably present, and investigation of the long-term variability in the spectra (broad lines and continuum) of this object can provide information about the accretion disk structure.

As mentioned in Paper I (and papers quoted in Sect. 1), the geometry of the BLR of 3C 390.3 probably has a disk-like structure.

From Table 4, we conclude that the distance (in velocity) between the blue and red peaks was smaller in 2002–2007, when the continuum flux was relatively high, while it was larger when the continuum flux was relatively weak. The similar anticorrelation in 1995–1999 was found in Paper I. This can be expected in the case where broad emission lines originate in the AD, i.e., are produced by changes in the dimension of the disk. The velocity difference increases, corresponding to smaller radii (emission from the part of the disk near to the BH), as the continuum flux decreases. When the continuum flux increases the velocity difference decreases, i.e., the contribution of disk parts farther from the BH is more significant.

If the BLR, however, consists of 2 regions - AD and sub-parcsec-jet (Arshakian et al. 2010), then, in the maximum activity state, most radiation is produced by the jet and the velocity difference between blue and red peak decreases (further from BH). Otherwise, in contrast (i.e., in the minimum state of activity) most radiation is caused by AD and the velocity difference increases (near BH). We also note that in blazars, where the emission is dominated by the extremely variable, non-thermal radiation from the jet, the few spectroscopic studies available (since lines are often absent for these objects) have shown that in some cases there is a weak (or no) correlation between the line and continuum emission. This result was also reported by Benitez et al. (2009) on the basis of their spectroscopical monitoring of blazar 3C 454.3 from September 2003 to July 2008. They found that Mg II emission lines respond proportionally to the continuum variations when the source is in a low-activity state. In contrast, close to the optical outbursts detected in 2005 and 2007, the Mg II emission lines showed little response to the continuum flux variations.

In any case, the profile evolution indicates that some additional emission is present, as seen in the first period 1997–1998 for the central part of the line, and at the end of the monitoring campaign, when a difference was measured for the red wings of Hβ and Hα lines. These results could be interpreted by some kind of two-component model (see also Popović et al. 2004; Bon et al. 2009) but may also caused by the perturbations in the disk (due to shock waves), providing support for a disk model (as noted in Paper I). A pure disk model would be consistent with the absence of a lag between flux variations in the red and blue wings of the lines, and the line shapes of Hα and Hβ (with the blue and red peaks) being indicative of the relativistic disk emission.

If one were to propose pure disk geometry (without additional emission), then the various bumps in line profiles may be caused by spiral shocks in the disk. As we have shown in Sect. 3.2, the Hβ and continuum light curves can be decomposed into three functions $F_A(t)$, $F_B(t)$, and $G(t)$. As discussed in Sect. 3.5, the time lag between $F_A^c(t)$ and $F_A^{H\beta}(t)$ is zero, while between $F_B^c(t)$ and $F_B^{H\beta}(t)$ it is around 600 days, which is close to the bump in CCF curves (see Fig. 17). The lag between the $G(t)$ functions corresponds to the lag between the continuum and line fluxes. This may indicate that we have a complex scenario in which the variability has two distinct periods. Assuming emission to originate in the disk, if a perturbation appears close to the black hole, the continuum flux may first increase, after which (at around 90-100 days) the line flux increases. This is well described by the lag in the function $G(t)$. On the other side, one may expect some of the continuum to be emitted from the line emission region (as e.g. due to free-bound transitions), which may be described by $F_A$ (lag zero between continuum and line). It is interesting that we have an additional lag of around 600 days in the flux ($F_B$), which may be caused by the perturbation seen in the flash, after which the perturbation spreads through the disk at super-sonic velocities.

The line profile variations and a possible model of the spiral shocks in the disk will be presented in Paper II of this series.

### 4.4. Disk versus a bipolar radial outflow

As noted in the Introduction, some models can explain the double-peaked broad emission line profiles in 3C 390.3. As aforementioned, a very popular model generates these lines in a relativistic disk. However, there is some contradiction between this model and the polarization observations of 3C 390.3. Smith et al. (2004, 2005) proposed a two-component scattering model to explain the polarization properties of the optical continuum and Hα broad emission lines in Sy1 nuclei. They hypothesized that both the compact equatorial scattering region located inside the torus and the polar scattering regions are present in all Seyfert galaxies, and that their polarization properties can be broadly understood in terms of an orientation effect. On the one hand, Smith at al. (2004) suggested that the broad emission lines originate in a rotating emission line disk surrounded by a coplanar scattering region. Thus, in polarization light a double-peaked profile of Hα will be observed, as in total light, and the degree of polarization will be maximal in the wings of the line and minimal in the line core. The position angle (PA) of polarization is aligned with the projected disk rotation axis and hence with the radio source axis (e.g. E-vector is parallel to the pc-jet axis). On the other hand, the polarization is caused by far-field scattering in the polar illumination cones of the circumnuclear torus (Smith et al. 2004). In polarization light, the Hα profile has only one central peak and the position angle (PA) of polarization is perpendicular to the projected disk rotation axis or pc-jet axis (e.g. E-vector is perpendicular to the pc-jet axis). From spectropolarimetric observations of 3C390.3 in Hα region taken at the 4.2 m Wiliam Herschel telescope in 1995 and 1997, Corbett et al. (1998, 2000) obtained the following main results:

a) The optical polarization E vector is very closely aligned with the radio source axis, as for the case of equatorial scattering region.

b) In polarized light, the Hα profile has a broad central peak centered close to the systematic velocity and is also slightly asymmetric, the blue wing being somewhat less extended than the red. This result favors polar scattering and is inconsistent with an equatorial scattering model because a double peak in the polarized line profile is not observed.

c) The P(λ)- the percentage polarization of the continuum changed significantly between 1995 and 1997, presumably because there is a variable, intrinsically polarized, synchrotron contribution to the continuum (from a pc-jet component).

Corbett at al. (1998, 2000) considered a model in which Hα photons emitted by a biconical flow are scattered by the inner wall of the torus. Thus, the scattering plane is perpendicular to the radio jet (e.g. E-vector parallel to pc-jet axis) and produces a single-peaked scattered Hα line profile. However, this model does not agree with the optical monitoring data, for which the following results are obtained:

1) The CCF-analysis for the Hβ and Hα line wings in 3C390 did not reveal any significant delay in the variations of the line wings with respect to the central part of the line, or relative to each other. The flux in the Hβ line wings and core also varied simultaneously. These results are indicative of predominantly circular motions in the BLR and favor a model in which the main contribution to the broad lines flux originates in the accretion disk and not the biconical flow (Dietrich et al. 1998; PaperI).

2) Livio & Xu (1997) demonstrated that the double-peaked lines seen in 3C390.3 cannot be produced by two line-emitting streams, since the emission region of the far cone jet would be obscured by the optically thick accretion disk. The aforementioned simple scattering models obviously would be unable to explain the available polarimetric and optic observations.

The BLR in 3C390.3 may have a complex structure and broad line fluxes that originate in several components, such as an accretion disk, a spherical ensemble of clouds following virilized orbits, a bipolar radial outflow, or pc-jet.

## 5. Conclusion

We have presented a long-term spectroscopic monitoring campaign of 3C 390.3. We have constructed the continuum and Hα and Hβ light curves, performed cross correlation between the continuum flux and line fluxes as well as between the line parts. In addition, we have analyzed any changes in Hβ and Hα line profiles and compiled mean and rms profiles. From our analysis and inspection of the observed data, we have reached the following conclusions:

(i) During 1995–2007, the Hα and Hβ broad emission lines and continuum varied 3-5 times. The amplitude of variability F(var) was larger in the continuum (0.46) than in either Hβ (0.38) or Hα (0.35). The blue Hβ and Hα wings were always brighter and vary slightly more strongly than those in the red. The Hα and Hβ profiles varied in a way similar to those of Sy 1 type in the maximum activity state and Sy1.8 type in the minimum activity state.

(ii) We found that QPOs exist in the Hβ and continuum light curve, as commonly observed for stellar-mass black holes. The periodicity of the light curves is probably connected with some shock waves near the supermassive black hole spreading in the outer part of the disk, but on the other hand, we cannot exclude the contribution of either ejection or jets to QPOs. Our results imply that the QPO variations in the observed flux ratio of the blue to red Hβ wings, of period ~10 years (P~10 yr), probably exist, as previously detected by Veilleux and Zheng (1991),.

(iii) Cross-correlation of the continuum and Hβ broad emission line fluxes shows a lag ~95 days, as in both Paper I (in 1998–1999, ~100 days) and Sergeev et al. (2002, lag~89 days). There is no significant lag between both the blue and red wings and the core, which is indicative of predominantly circular motions in the BLR of 3C390.3 and a disk-like geometry. For Hα, we found a lag of around 120 days. This difference in lags as well as in the FWHM of Hα and Hβ may imply that there is some stratification in the BLR (disk) of 3C 390.3.

(iv) The Hα and Hβ line fluxes and parts of lines are strongly correlated with the continuum flux, indicating that the ionizing continuum was a good extrapolation of the optical one.

(v) By comparing the Hβ and Hα line profiles, we found that the broad emission region has a disk-like structure, but there probably exists a non-disk component, or also a disk-like one of different parameters that contributes to the line emission. More details about the line shape variation and any possible disk structure will be given in a forthcoming paper.

## Acknowledgments

This work was supported by INTAS (grant N96-0328), RFBR (grants N97-02-17625 N00-02-16272, N03-02-17123, 06-02-16843, and N09-02-01136), State program 'Astronomy' (Russia), CONACYT research grant 39560-F and 54480 (México), PROMEP/103.5/08/4722 grant and the Ministry of Science and Technological Development of Republic of Serbia through the project Astrophysical Spectroscopy of Extragalactic Objects (146002). L. Č. P., W. K. and D. I. are grateful to the Alexander von Humboldt foundation for support in the frame of program "Research Group Linkage". We would like to thank to the anonymous referee for very useful comments.

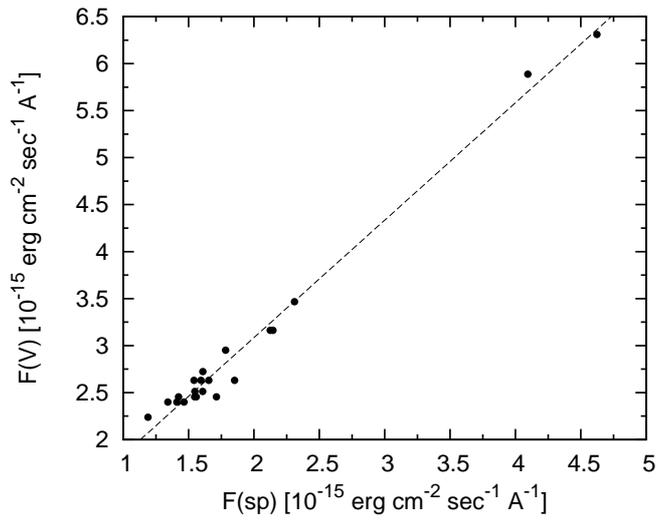

**Fig. 1.** The correlation between the photometric – F(V) and spectral – F(sp) fluxes.

**Table 1.** The log of spectroscopic observations.

| N | UT-date | JD 2400000+ | CODE* | Aperture arcsec | Sp.range Å | Res Å | Seeing arcsec |
|---|---------|-------------|-------|-----------------|------------|-------|---------------|
| 1 | 2 | 3 | 4 | 5 | 6 | 7 | 8 |
| 1 | 1995Apr25 | 49832.424 | L(G) | 3.0×3.6 | 4000-5400 | - | 2 |
| 2 | 1995May26 | 49863.375 | L(T) | 3.6×3.6 | 4900-5450 | 4.5 | 2 |
| 3 | 1995Nov17 | 50039.156 | L(G) | 3.6×8.4 | 3700-5800 | - | - |
| 4 | 1995Nov29 | 50051.143 | L(G) | 2.0×8.4 | 4600-5400 | - | - |
| 5 | 1995Nov30 | 50052.149 | L(G) | 2.0×10.8 | 4700-5600 | - | - |
| 6 | 1996Feb14 | 50127.602 | L(N) | 1.5×6.0 | 3100-5800 | 6 | 2 |
| 7 | 1996Mar20 | 50162.580 | L(N) | 2.0×6.0 | 3600-5600 | 6 | 3 |
| 8 | 1996Mar20 | 50162.587 | L(N) | 2.0×6.0 | 4700-7400 | 6 | 3 |
| 9 | 1996Mar21 | 50163.553 | L(N) | 2.0×6.0 | 3600-5600 | 6 | 3 |
| 10 | 1996Mar21 | 50163.572 | L(N) | 2.0×6.0 | 4700-7400 | 6 | 3 |
| 11 | 1996Jun15 | 50249.542 | L(G) | 2.0×6.0 | 4600-5500 | 5 | 2.5 |
| 12 | 1996Jul12 | 50276.567 | L(G) | 2.0×6.0 | 4600-5500 | 5 | 1.4 |
| 13 | 1996Jul12 | 50276.570 | L(G) | 2.0×6.0 | 6370-7270 | 5 | 1.4 |
| 14 | 1996Jul13 | 50277.556 | L(G) | 2.0×6.0 | 4600-5500 | 5 | 1.6 |
| 15 | 1996Jul17 | 50281.434 | L(G) | 2.0×6.0 | 4100-5800 | 8 | 1.3 |
| 16 | 1996Aug10 | 50305.489 | L(G) | 2.0×6.0 | 4600-5575 | 5 | 3.2 |
| 17 | 1996Sep11 | 50338.800 | Z2 | 4.2×13.8 | 3850-5725 | 5 | 5 |
| 18 | 1996Nov03 | 50390.431 | L(G) | 2.0×6.0 | 4600-5500 | 4 | 3 |
| 19 | 1996Nov05 | 50392.621 | L(G) | 2.0×6.0 | 4621-5521 | 4 | 3-4 |
| 20 | 1997Mar04 | 50511.622 | L(G) | 2.0×6.0 | 4650-5550 | 5 | 1.8 |
| 21 | 1997May30 | 50599.370 | L(G) | 2.0×6.0 | 4700-5600 | 5 | 2.5 |
| 22 | 1997Jul26 | 50656.499 | L(G) | 2.0×6.0 | 4700-5675 | 5 | 3-4 |
| 23 | 1997Aug30 | 50691.463 | L(G) | 2.0×6.0 | 4700-5600 | 5 | 2 |
| 24 | 1997Sep10 | 50701.576 | L(N) | 2.0×6.0 | 3900-6600 | 6 | - |
| 25 | 1997Sep10 | 50701.566 | L(N) | 2.0×6.0 | 5200-7900 | 6 | - |
| 26 | 1997Dec26 | 50808.582 | L(U) | 2.0×6.0 | 4800-5700 | 4 | 3.0 |
| 27 | 1997Dec27 | 50810.184 | L(U) | 2.0×6.0 | 4750-5650 | 4 | 4.0 |
| 28 | 1997Dec27 | 50810.203 | L(U) | 2.0×6.0 | 6400-7300 | 4 | 4.0 |
| 29 | 1997Dec28 | 50811.194 | L(U) | 2.0×6.0 | 4750-5650 | 4 | 2.6 |
| 30 | 1997Dec28 | 50811.206 | L(U) | 2.0×6.0 | 6400-7300 | 4 | 2.6 |
| 31 | 1997Dec30 | 50813.195 | L(G) | 2.0×6.0 | 3600-7200 | 12 | 3.2 |
| 32 | 1998Jan22 | 50835.631 | L(N) | 2.0×6.0 | 3800-6200 | 6 | 2 |
| 33 | 1998Feb23 | 50867.560 | L(N) | 2.0×6.0 | 3800-6200 | 6 | 2 |
| 34 | 1998Feb23 | 867.573 | L(N) | 2.0×6.0 | 5000-7400 | 6 | 2 |
| 35 | 1998May06 | 50940.354 | L(N) | 2.0×6.0 | 3740-6200 | 6 | 3 |
| 36 | 1998Jun25 | 50990.302 | L(N) | 2.0×6.0 | 3600-6100 | 6 | 3 |
| 37 | 1998Jul16 | 51010.719 | GHO | 2.5×6.0 | 4200-7400 | 15 | 2 |
| 38 | 1998Jul25 | 51019.723 | GHO | 2.5×6.0 | 4440-7640 | 15 | 2.2 |
| 39 | 1998Jul28 | 51023.491 | L(G) | 2.0×6.0 | 4800-5700 | 5 | 1.3 |
| 40 | 1998Jul30 | 51025.431 | L(G) | 2.0×6.0 | 4800-5700 | 5 | 2 |
| 41 | 1998Aug30 | 51055.551 | L(N) | 2.0×6.0 | 4000-6500 | 6 | 1.8 |
| 42 | 1998Aug30 | 51055.562 | L(N) | 2.0×6.0 | 5250-7750 | 6 | 1.7 |
| 43 | 1998Sep19 | 51076.004 | Z1 | 4.2×19.8 | 4100-5900 | 8 | 2.0 |
| 44 | 1998Sep25 | 51081.429 | GHO | 2.5×6.0 | 4230-7430 | 15 | 2 |
| 45 | 1998Sep26 | 51082.429 | GHO | 2.5×6.0 | 4210-7410 | 15 | 2 |
| 46 | 1998Sep27 | 51083.429 | GHO | 2.5×6.0 | 4210-7410 | 15 | 2 |
| 47 | 1998Oct25 | 51112.259 | L(G) | 2.0×6.0 | 4700-5600 | 5 | 3 |
| 48 | 1998Dec02 | 51149.625 | GHO | 2.5×6.0 | 4220-7600 | 15 | - |
| 49 | 1999Jul13 | 51372.516 | L(P) | 2.0×6.0 | 4420-5655 | 4 | 1.5 |
| 50 | 1999Aug19 | 51410.309 | L(P) | 2.0×6.0 | 4500-5740 | 4 | 2 |
| 51 | 1999Sep04 | 51425.904 | L(U) | 2.0×6.0 | 3600-6024 | 7 | 1.5 |
| 52 | 1999Oct03 | 51454.674 | GHO | 2.5×6.0 | 4060-7330 | 15 | 2 |
| 53 | 1999Sep05 | 51427.272 | L(P) | 2.0×6.0 | 4900-7320 | 7 | 1.5 |
| 54 | 1999Oct04 | 51455.664 | L(U) | 2.0×6.0 | 4480-5728 | 4 | 1.3 |
| 55 | 1999Oct09 | 51461.531 | L(N) | 2.0×6.0 | 4250-6580 | 8 | 3 |
| 56 | 2000Jul19 | 51745.365 | L(U) | 2.0×6.0 | 4390-5590 | 5 | 2.2 |
| 57 | 2000Jul22 | 51748.232 | L(U) | 2.0×6.0 | 4390-5580 | 5 | - |
| 58 | 2000Jul30 | 51756.293 | L(U) | 2.0×6.0 | 3690-6090 | 8 | 1.6 |

**Table 1.** Continued.

| N | UT-date | JD 2400000+ | CODE* | Aperture arcsec | Sp.range Å | Res Å | Seeing arcsec |
|---|---------|-------------|-------|-----------------|-----------|-------|---------------|
| 1 | 2 | 3 | 4 | 5 | 6 | 7 | 8 |
| 59 | 2000Jul30 | 51756.317 | L(U) | 2.0×6.0 | 5640-8040 | 8 | 1.6 |
| 60 | 2000Oct05 | 51823.142 | L(U) | 2.0×6.0 | 3690-6030 | 8 | 1.2 |
| 61 | 2000Nov18 | 51867.136 | L(U) | 2.0×6.0 | 4490-5690 | 5 | 1.0 |
| 62 | 2000Nov19 | 51868.128 | L(U) | 2.0×6.0 | 4490-5690 | 5 | 1.2 |
| 63 | 2001Jan16 | 51925.631 | L(U) | 2.0×6.0 | 3690-6040 | 8 | 1.8 |
| 64 | 2001Jan16 | 51925.647 | L(U) | 2.0×6.0 | 5690-8040 | 8 | 1.8 |
| 65 | 2001Jan20 | 51929.623 | L(U) | 2.0×6.0 | 3690-6040 | 8 | 1.8 |
| 66 | 2001Jan20 | 51929.639 | L(U) | 2.0×6.0 | 5740-7990 | 8 | 1.8 |
| 67 | 2001Mar13 | 51981.950 | GHO | 2.5×6.0 | 4116-7436 | 15 | 2.0 |
| 68 | 2001May05 | 52034.660 | GHO | 2.5×6.0 | 4160-7470 | 15 | - |
| 69 | 2001May14 | 52043.900 | GHO | 2.5×6.0 | 3980-7280 | 15 | - |
| 70 | 2001Jun14 | 52074.850 | GHO | 2.5×6.0 | 4010-7320 | 15 | - |
| 71 | 2001Jun14.5 | 52075.350 | GHO | 2.5×6.0 | 4010-7320 | 15 | - |
| 72 | 2001Jun15 | 52075.850 | GHO | 2.5×6.0 | 4010-7320 | 15 | - |
| 73 | 2001Oct10 | 52192.600 | GHO | 2.5×6.0 | 4030-7336 | 15 | 2.3 |
| 74 | 2001Nov23 | 52237.272 | L(U) | 2.0×6.0 | 3620-6000 | 8 | 3.5 |
| 75 | 2002Feb21 | 52327.200 | L(U) | 2.0×6.0 | 5690-8040 | 8 | 2.5 |
| 76 | 2002Feb21 | 52327.226 | L(U) | 2.0×6.0 | 3490-5840 | 8 | 2.5 |
| 77 | 2002Mar05 | 52339.019 | GHO | 2.5×6.0 | 3880-7190 | 15 | 2.0 |
| 78 | 2002Apr03 | 52368.005 | GHO | 2.5×6.0 | 4290-5960 | 8 | - |
| 79 | 2002Jun03 | 52428.968 | GHO | 2.5×6.0 | 5740-7430 | 8 | - |
| 80 | 2002Jun04 | 52429.950 | GHO | 2.5×6.0 | 3980-7300 | 15 | - |
| 81 | 2002Jun05 | 52430.938 | GHO | 2.5×6.0 | 4240-5910 | 8 | - |
| 82 | 2002Jun24 | 52450.385 | L(U) | 2.0×6.0 | 3500-5840 | 8 | 5 |
| 83 | 2002Jun24 | 52450.410 | L(U) | 2.0×6.0 | 5640-7990 | 8 | 5 |
| 84 | 2002Jul09 | 52464.444 | L(U) | 2.0×6.0 | 3500-5840 | 8 | 2 |
| 85 | 2002Jul13 | 52469.408 | L(U) | 2.0×6.0 | 3500-5840 | 8 | 2 |
| 86 | 2002Jul13 | 52469.433 | L(U) | 2.0×6.0 | 5640-7990 | 8 | 2 |
| 87 | 2002Aug08 | 52495.272 | L(U) | 2.0×6.0 | 4290-5450 | 8 | - |
| 88 | 2002Aug16 | 52502.795 | GHO | 2.5×6.0 | 4250-5930 | 8 | - |
| 89 | 2002Oct14 | 52562.318 | L(U) | 2.0×6.0 | 5640-7990 | 8 | 2 |
| 90 | 2002Oct14 | 52562.342 | L(U) | 2.0×6.0 | 3500-5840 | 8 | 2 |
| 91 | 2002Nov13 | 52591.659 | GHO | 2.5×6.0 | 5720-7410 | 8 | - |
| 92 | 2002Nov14 | 52592.659 | GHO | 2.5×6.0 | 5720-7410 | 8 | - |
| 93 | 2002Nov15 | 52593.671 | GHO | 2.5×6.0 | 3800-7115 | 15 | - |
| 94 | 2002Dec11 | 52620.389 | L(U) | 2.0×6.0 | 3650-5990 | 8 | 2 |
| 95 | 2003May08 | 52768.217 | L(U) | 2.0×6.0 | 3700-6040 | 8 | 1.5 |
| 96 | 2003May10 | 52770.219 | L(U) | 2.0×6.0 | 5750-8090 | 8 | 1.5 |
| 97 | 2003May23 | 52782.966 | GHO | 2.5×6.0 | 3540-7165 | 15 | - |
| 98 | 2003May24 | 52783.976 | GHO | 2.5×6.0 | 4220-6060 | 8 | - |
| 99 | 2003May25 | 52784.973 | GHO | 2.5×6.0 | 5580-7440 | 8 | - |
| 100 | 2003May26 | 52785.979 | GHO | 2.5×6.0 | 4230-6040 | 8 | - |
| 101 | 2003Jun22 | 52812.963 | GHO | 2.5×6.0 | 4270-5980 | 8 | - |
| 102 | 2003Sep18 | 52900.794 | GHO | 2.5×6.0 | 3710-7100 | 15 | - |
| 103 | 2003Sep20 | 52902.803 | GHO | 2.5×6.0 | 5630-7340 | 8 | - |
| 104 | 2003Oct21 | 52933.672 | GHO | 2.5×6.0 | 3840-7220 | 15 | - |
| 105 | 2003Nov19 | 52962.637 | GHO | 2.5×6.0 | 3750-7132 | 15 | - |
| 106 | 2003Nov20 | 52963.608 | GHO | 2.5×6.0 | 4240-5950 | 8 | - |
| 107 | 2004Mar01 | 53066.358 | L(U) | 2.0×6.0 | 5800-8100 | 8 | 2 |
| 108 | 2004Mar01 | 53065.507 | L(U) | 2.0×6.0 | 3740-6090 | 8 | 2 |
| 109 | 2004May20 | 53145.982 | GHO | 2.5×6.0 | 4191-5910 | 8 | - |
| 110 | 2004Jun12 | 53168.955 | GHO | 2.5×6.0 | 4210-5920 | 8 | - |
| 111 | 2004Jun13 | 53169.966 | GHO | 2.5×6.0 | 5570-7320 | 8 | - |
| 112 | 2004Jun15 | 53171.944 | GHO | 2.5×6.0 | 3760-7148 | 15 | - |
| 113 | 2004Aug19 | 53236.808 | GHO | 2.5×6.0 | 4280-5940 | 8 | - |
| 114 | 2004Aug20 | 53237.821 | GHO | 2.5×6.0 | 5630-7360 | 8 | - |
| 115 | 2004Aug21 | 53238.858 | GHO | 2.5×6.0 | 3700-7093 | 15 | - |
| 116 | 2004Sep06 | 53254.771 | GHO | 2.5×6.0 | 4190-5900 | 8 | - |
| 117 | 2004Dec18 | 53358.140 | L(S) | 1.0×6.1 | 3100-7300 | 10 | 4.5 |

**Table 1.** Continued.

| N | UT-date | JD 2400000+ | CODE* | Aperture arcsec | Sp.range Å | Res Å | Seeing arcsec |
|---|---------|-------------|-------|-----------------|------------|-------|---------------|
| 1 | 2 | 3 | 4 | 5 | 6 | 7 | 8 |
| 118 | 2004Dec19 | 53359.130 | L(S) | 1.0×6.1 | 3100-7300 | 10 | 2.0 |
| 119 | 2004Dec20 | 53360.150 | L(S) | 1.0×6.1 | 3100-7300 | 10 | 1.6 |
| 120 | 2005Apr15 | 53475.997 | GHO | 2.5×6.0 | 4250-5960 | 8 | - |
| 121 | 2005Apr16 | 53476.993 | GHO | 2.5×6.0 | 5540-7250 | 8 | - |
| 122 | 2005Apr18 | 53478.992 | GHO | 2.5×6.0 | 3790-7173 | 15 | - |
| 123 | 2005May13 | 53503.972 | GHO | 2.5×6.0 | 4220-5910 | 8 | - |
| 124 | 2005Jun10 | 53531.925 | GHO | 2.5×6.0 | 4280-5970 | 8 | - |
| 125 | 2005Aug30 | 53612.758 | GHO | 2.5×6.0 | 4330-5990 | 8 | - |
| 126 | 2005Aug31 | 53613.773 | GHO | 2.5×6.0 | 4330-6000 | 8 | - |
| 127 | 2005Sep01 | 53614.750 | GHO | 2.5×6.0 | 4330-6000 | 8 | - |
| 128 | 2005Dec06 | 53711.420 | L(S) | 1.0×6.1 | 3100-7300 | 10 | 4.0 |
| 129 | 2006Jun04 | 53891.490 | L(S) | 1.0×6.1 | 3100-7300 | 10 | 1.5 |
| 130 | 2006Sep01 | 53979.790 | GHO | 2.5×6.0 | 3500-7044 | 15 | 2.1 |
| 131 | 2006Sep16 | 53994.780 | GHO | 2.5×6.0 | 3530-7070 | 15 | 3.4 |
| 132 | 2006Sep17 | 53995.750 | GHO | 2.5×6.0 | 4150-5950 | 7.5 | 2.4 |
| 133 | 2006Sep18 | 53996.700 | GHO | 2.5×6.0 | 3470-7020 | 15 | 2.3 |
| 134 | 2006Sep19 | 53997.740 | GHO | 2.5×6.0 | 4120-5920 | 7.5 | 2.4 |
| 135 | 2006Sep20 | 53998.750 | GHO | 2.5×6.0 | 3470-7020 | 15 | 2.7 |
| 136 | 2006Oct28 | 54036.630 | GHO | 2.5×6.0 | 3710-7270 | 15 | 2.7 |
| 137 | 2006Oct29 | 54037.640 | GHO | 2.5×6.0 | 4230-6030 | 7.5 | 2.2 |
| 138 | 2006Nov01 | 54040.650 | GHO | 2.5×6.0 | 4158-5958 | 7.5 | 2.3 |
| 139 | 2006Dec18 | 54088.210 | L(S) | 1.0×6.1 | 3100-7300 | 10 | 2.0 |
| 140 | 2007Jan12 | 54112.640 | L(S) | 1.0×6.1 | 3100-7300 | 10 | 2.9 |
| 141 | 2007Jan13 | 54113.650 | L(S) | 1.0×6.1 | 3100-7300 | 10 | 2.2 |
| 142 | 2007Feb12 | 54143.610 | L(S) | 1.0×6.1 | 3100-7300 | 10 | 2.3 |
| 143 | 2007May22 | 54242.890 | GHO | 2.5×6.0 | 3745-7335 | 15 | 3.7 |
| 144 | 2007May23 | 54243.880 | GHO | 2.5×6.0 | 4185-5985 | 7.5 | 4.2 |
| 145 | 2007Jun20 | 54271.870 | GHO | 2.5×6.0 | 3700-7150 | 15 | 2.0 |
| 146 | 2007Jun21 | 54272.960 | GHO | 2.5×6.0 | 4282-6090 | 7.5 | 2.7 |
| 147 | 2007Aug09 | 54321.820 | GHO | 2.5×6.0 | 3860-7430 | 15 | 2.3 |
| 148 | 2007Aug11 | 54323.800 | GHO | 2.5×6.0 | 3860-7430 | 15 | 6.6 |
| 149 | 2007Aug12 | 54324.810 | GHO | 2.5×6.0 | 4330-6140 | 7.5 | 3.2 |
| 150 | 2007Sep03 | 54346.730 | GHO | 2.5×6.0 | 3830-7417 | 15 | 3.4 |
| 151 | 2007Sep04 | 54347.720 | GHO | 2.5×6.0 | 4320-6130 | 7.5 | 3.6 |
| 152 | 2007Sep05 | 54348.720 | GHO | 2.5×6.0 | 4140-5950 | 7.5 | 2.6 |
| 153 | 2007Oct16 | 54389.690 | GHO | 2.5×6.0 | 3880-7440 | 15 | 2.1 |
| 154 | 2007Oct18 | 54391.630 | GHO | 2.5×6.0 | 4200-6000 | 7.5 | 2.5 |
| 155 | 2007Nov02 | 54406.650 | GHO | 2.5×6.0 | 4180-5990 | 7.5 | 3.0 |
| 156 | 2007Nov04 | 54408.610 | GHO | 2.5×6.0 | 3810-7380 | 15 | 2.9 |
| 157 | 2007Nov08 | 54412.690 | GHO | 2.5×6.0 | 3820-7390 | 15 | 2.8 |

**Notes.** – Col.(1): Number. Col.(2): UT date. Col.(3): Julian date (JD). Col.(4): CODE* . Col.(5): Projected spectrograph entrance apertures. Col.(6): Wavelength range covered. Col.(7): Spectral resolution. Col.(8): Mean seeing in arcsec.
[*] L(G,T,U,P,S) - 6m telescope in prime focus with different variants of the long slit spectrographs and CCD; L(N) - 6m telescope, Nasmyth focus; Z(1,2) - 1 m Zeiss, Cassegrain focus; GHO - 2.1 m telescope GHO (Mexico), Cassegrain focus and B&Ch spectrograph.

**Table 2.** Observed continuum, the broad Hα and Hβ fluxes, reduced to the 6 m telescope aperture 2″ × 6″.

| N | UT-date | JD 2400000+ | F(5100)±σ 10⁻¹⁵ erg cm⁻² s⁻¹Å⁻¹ | F(Hα)±σ 10⁻¹³ erg cm⁻² s⁻¹ | F(Hβ)±σ 10⁻¹³ erg cm⁻² s⁻¹ |
|---|---|---|---|---|---|
| 1 | 2 | 3 | 4 | 5 | 6 |
| 1 | 1995Apr25 | 49832.424 | 1.37±0.04 | | 1.67±0.07 |
| 2 | 1995May26 | 49863.375 | 1.55±0.05 | | 2.02±0.08 |
| 3 | 1995Nov17 | 50039.156 | 2.46±0.07 | | 2.18±0.09 |
| 4 | 1995Nov29 | 50051.143 | 2.47±0.07 | | 2.27±0.10 |
| 5 | 1995Nov30 | 50052.149 | 2.55±0.08 | | 1.89±0.08 |
| 6 | 1996Feb14 | 50127.602 | 2.27±0.07 | | 2.73±0.11 |
| 7 | 1996Mar20 | 50162.580 | 2.49±0.07 | | 2.83±0.12 |
| 8 | 1996Mar20 | 50162.587 | | 13.90±1.17 | |
| 9 | 1996Mar21 | 50163.553 | 2.52±0.08 | | 3.15±0.13 |
| 10 | 1996Mar21 | 50163.572 | | 11.19±0.94 | |
| 11 | 1996Jun15 | 50249.542 | 1.74±0.05 | | 2.45±0.10 |
| 12 | 1996Jul12 | 50276.567 | 1.49±0.04 | | 2.68±0.11 |
| 13 | 1996Jul12 | 50276.570 | | 12.13±1.02 | |
| 14 | 1996Jul13 | 50277.556 | 1.48±0.04 | | 2.44±0.10 |
| 15 | 1996Jul17 | 50281.434 | 1.57±0.05 | | 2.38±0.10 |
| 16 | 1996Aug10 | 50305.489 | 1.72±0.05 | | 2.47±0.10 |
| 17 | 1996Sep11 | 50338.800 | 1.75±0.05 | | 1.91±0.08 |
| 18 | 1996Nov03 | 50390.431 | 1.55±0.05 | | 2.02±0.09 |
| 19 | 1996Nov05 | 50392.621 | 1.50±0.05 | | 1.84±0.08 |
| 20 | 1997Mar04 | 50511.622 | 1.10±0.03 | | 1.28±0.05 |
| 21 | 1997May30 | 50599.370 | 1.19±0.04 | | 1.42±0.06 |
| 22 | 1997Jul26 | 50656.499 | 0.97±0.03 | | 1.30±0.05 |
| 23 | 1997Aug30 | 50691.463 | 1.11±0.03 | | 1.28±0.05 |
| 24 | 1997Sep10 | 50701.566 | | 7.28±0.61 | |
| 25 | 1997Sep10 | 50701.576 | 1.21±0.04 | | 1.12±0.05 |
| 26 | 1997Dec26 | 50808.582 | 1.68±0.05 | | 1.53±0.06 |
| 27 | 1997Dec27 | 50810.184 | 1.56±0.05 | | 1.79±0.08 |
| 28 | 1997Dec27 | 50810.203 | | 9.74±0.82 | |
| 29 | 1997Dec28 | 50811.194 | 1.49±0.04 | | 1.56±0.07 |
| 30 | 1997Dec28 | 50811.206 | | 7.99±0.67 | |
| 31 | 1997Dec30 | 50813.195 | 1.58±0.05 | | 1.45±0.06 |
| 32 | 1998Jan22 | 50835.631 | 1.81±0.05 | | 1.61±0.07 |
| 33 | 1998Feb23 | 50867.560 | 1.62±0.05 | | 1.87±0.08 |
| 34 | 1998Feb23 | 50867.573 | | 6.06±0.51 | |
| 35 | 1998May06 | 50940.354 | 1.87±0.06 | | 2.29±0.10 |
| 36 | 1998Jun25 | 50990.302 | 1.78±0.05 | | 2.21±0.09 |
| 37 | 1998Jul16 | 51010.719 | 1.56±0.05 | 8.12±0.68 | 2.35±0.10 |
| 38 | 1998Jul25 | 51019.723 | 1.43±0.04 | | 2.34±0.10 |
| 39 | 1998Jul28 | 51023.491 | 1.55±0.05 | | 2.14±0.09 |
| 40 | 1998Jul30 | 51025.431 | 1.40±0.04 | | 1.95±0.08 |
| 41 | 1998Aug30 | 51055.551 | 1.45±0.04 | | 2.04±0.09 |
| 42 | 1998Aug30 | 51055.562 | | 10.30±0.87 | |
| 43 | 1998Sep19 | 51076.004 | 1.65±0.05 | | 2.07±0.09 |
| 44 | 1998Sep25 | 51081.429 | 1.51±0.05 | | 2.10±0.09 |
| 45 | 1998Sep26 | 51082.429 | 1.59±0.05 | | 2.14±0.09 |
| 46 | 1998Sep27 | 51083.429 | 1.55±0.05 | 6.70±0.56 | 2.20±0.09 |
| 47 | 1998Oct25 | 51112.259 | 1.34±0.04 | | 2.11±0.09 |
| 48 | 1998Dec02 | 51149.625 | 1.45±0.04 | 7.68±0.65 | 2.15±0.09 |
| 49 | 1999Jul13 | 51372.516 | 1.19±0.04 | | 1.37±0.06 |
| 50 | 1999Aug19 | 51410.309 | 1.34±0.04 | | 1.61±0.07 |
| 51 | 1999Sep04 | 51425.904 | 1.43±0.04 | | 1.65±0.07 |
| 52 | 1999Sep05 | 51427.272 | | 7.40±0.62 | |
| 53 | 1999Oct03 | 51454.674 | 1.62±0.05 | | 1.89±0.08 |
| 54 | 1999Oct04 | 51455.664 | 1.38±0.04 | | 1.80±0.08 |
| 55 | 1999Oct09 | 51461.531 | 1.55±0.05 | | 1.65±0.07 |
| 56 | 2000Jul19 | 51745.365 | 1.56±0.05 | | 1.92±0.08 |
| 57 | 2000Jul22 | 51748.232 | 1.60±0.05 | | 1.56±0.07 |
| 58 | 2000Jul30 | 51756.293 | 1.54±0.05 | | 1.98±0.08 |

**Table 2.** Continued.

| N | UT-date | JD 2400000+ | $F(5100)\pm\sigma$ $10^{-15}$ erg cm$^{-2}$ s$^{-1}$Å$^{-1}$ | $F(H\alpha)\pm\sigma$ $10^{-13}$ erg cm$^{-2}$ s$^{-1}$ | $F(H\beta)\pm\sigma$ $10^{-13}$ erg cm$^{-2}$ s$^{-1}$ |
|---|---|---|---|---|---|
| 1 | 2 | 3 | 4 | 5 | 6 |
| 59 | 2000Jul30 | 51756.317 | | 9.49±0.80 | |
| 60 | 2000Oct05 | 51823.142 | 1.41±0.04 | | 1.72±0.07 |
| 61 | 2000Nov18 | 51867.136 | 1.56±0.05 | | 1.82±0.08 |
| 62 | 2000Nov19 | 51868.128 | 1.66±0.05 | | 1.93±0.08 |
| 63 | 2001Jan16 | 51925.631 | 1.53±0.05 | | 1.93±0.08 |
| 64 | 2001Jan16 | 51925.647 | | 7.78±0.65 | |
| 65 | 2001Jan20 | 51929.623 | 1.47±0.04 | | 1.81±0.08 |
| 66 | 2001Jan20 | 51929.639 | | 9.59±0.81 | |
| 67 | 2001Mar13 | 51981.950 | 1.40±0.04 | | 1.88±0.08 |
| 68 | 2001May05 | 52034.660 | 1.39±0.04 | 7.38±0.62 | 1.56±0.07 |
| 69 | 2001May14 | 52043.900 | 1.35±0.04 | 6.85±0.58 | 1.56±0.07 |
| 70 | 2001Jun14 | 52074.850 | 1.34±0.04 | | 1.59±0.07 |
| 71 | 2001Jun14.5 | 52075.350 | | 6.77±0.57 | |
| 72 | 2001Jun15 | 52075.850 | 1.30±0.04 | | 1.55±0.07 |
| 73 | 2001Oct10 | 52192.600 | 1.08±0.03 | 6.25±0.52 | 1.22±0.05 |
| 74 | 2001Nov23 | 52237.272 | 1.37±0.04 | | 1.06±0.04 |
| 75 | 2002Feb21 | 52327.226 | 1.36±0.04 | | 1.09±0.05 |
| 76 | 2002Feb21 | 52327.200 | | 6.63±0.56 | |
| 77 | 2002Mar05 | 52339.019 | 1.20±0.04 | 5.60±0.47 | 0.94±0.04 |
| 78 | 2002Apr03 | 52368.005 | 1.20±0.04 | | 1.02±0.04 |
| 79 | 2002Jun03 | 52428.968 | | 6.60±0.55 | |
| 80 | 2002Jun04 | 52429.950 | | 6.47±0.54 | |
| 81 | 2002Jun05 | 52430.938 | 1.42±0.04 | | 1.21±0.05 |
| 82 | 2002Jun24 | 52450.385 | 1.72±0.05 | | 1.71±0.07 |
| 83 | 2002Jun24 | 52450.410 | | 5.86±0.49 | |
| 84 | 2002Jul09 | 52464.444 | 1.60±0.05 | | 1.54±0.06 |
| 85 | 2002Jul13 | 52469.408 | 1.85±0.06 | | 1.72±0.07 |
| 86 | 2002Jul13 | 52469.433 | | 5.93±0.50 | |
| 87 | 2002Aug08 | 52495.272 | 1.71±0.05 | | 1.42±0.06 |
| 88 | 2002Aug16 | 52502.795 | 1.61±0.05 | | 1.60±0.07 |
| 89 | 2002Oct14 | 52562.318 | | 9.03±0.76 | |
| 90 | 2002Oct14 | 52562.342 | 1.55±0.05 | | 1.82±0.08 |
| 91 | 2002Nov13 | 52591.659 | | 7.87±0.66 | |
| 92 | 2002Nov14 | 52592.659 | | 6.32±0.53 | |
| 93 | 2002Nov15 | 52593.671 | 1.36±0.04 | | 1.60±0.07 |
| 94 | 2002Dec11 | 52620.389 | 1.71±0.05 | | 1.60±0.07 |
| 95 | 2003May08 | 52768.217 | 1.96±0.06 | | 1.69±0.07 |
| 96 | 2003May10 | 52770.219 | | 6.40±0.54 | |
| 97 | 2003May23 | 52782.966 | 2.10±0.06 | | 1.69±0.07 |
| 98 | 2003May24 | 52783.976 | 2.15±0.06 | | 1.76±0.07 |
| 99 | 2003May25 | 52784.973 | | 6.68±0.56 | |
| 100 | 2003May26 | 52785.979 | 2.15±0.06 | | 1.68±0.07 |
| 101 | 2003Jun22 | 52812.963 | 2.28±0.07 | | 1.78±0.07 |
| 102 | 2003Sep18 | 52900.794 | | | 1.96±0.08 |
| 103 | 2003Sep20 | 52902.803 | | 7.78±0.65 | |
| 104 | 2003Oct21 | 52933.672 | 1.96±0.06 | 7.11±0.60 | 1.77±0.07 |
| 105 | 2003Nov19 | 52962.637 | 2.83±0.08 | | 1.99±0.08 |
| 106 | 2003Nov20 | 52963.608 | 2.93±0.09 | | 2.15±0.09 |
| 107 | 2004Mar01 | 53066.358 | | 9.15±0.77 | |
| 108 | 2004Mar01 | 53065.507 | 2.48±0.07 | | 2.25±0.09 |
| 109 | 2004May20 | 53145.982 | 2.01±0.06 | | 2.40±0.10 |
| 110 | 2004Jun12 | 53168.955 | 1.95±0.06 | | 2.32±0.10 |
| 111 | 2004Jun13 | 53169.966 | | 8.71±0.73 | |
| 112 | 2004Jun15 | 53171.944 | 1.95±0.06 | | 2.04±0.09 |
| 113 | 2004Aug19 | 53236.808 | 2.31±0.07 | | 2.11±0.09 |
| 114 | 2004Aug20 | 53237.821 | | 9.49±0.80 | |
| 115 | 2004Aug21 | 53238.858 | 2.31±0.07 | | 2.16±0.09 |
| 116 | 2004Sep06 | 53254.771 | 2.13±0.06 | | 2.16±0.09 |
| 117 | 2004Dec18 | 53358.140 | 3.13±0.09 | | 2.78±0.12 |

**Table 2.** Continued.

| N | UT-date | JD 2400000+ | $F(5100)\pm\sigma$ $10^{-15}$ erg cm$^{-2}$ s$^{-1}$Å$^{-1}$ | $F(H\alpha)\pm\sigma$ $10^{-13}$ erg cm$^{-2}$ s$^{-1}$ | $F(H\beta)\pm\sigma$ $10^{-13}$ erg cm$^{-2}$ s$^{-1}$ |
|---|---------|-------------|------|------|------|
| 1 | 2 | 3 | 4 | 5 | 6 |
| 118 | 2004Dec19 | 53359.130 | 2.92±0.09 | 10.69±0.90 | 2.67±0.11 |
| 119 | 2004Dec20 | 53360.150 | 2.87±0.09 | 11.23±0.94 | 2.76±0.12 |
| 120 | 2005Apr15 | 53475.997 | 3.19±0.1 | | 2.71±0.11 |
| 121 | 2005Apr16 | 53476.993 | | 9.69±0.81 | |
| 122 | 2005Apr18 | 53478.992 | 3.28±0.1 | 10.11±0.85 | 2.99±0.13 |
| 123 | 2005May13 | 53503.972 | 3.66±0.11 | | 2.92±0.12 |
| 124 | 2005Jun10 | 53531.925 | 4.25±0.13 | | 2.91±0.12 |
| 125 | 2005Aug30 | 53612.758 | 4.57±0.14 | | 3.50±0.15 |
| 126 | 2005Aug31 | 53613.773 | 4.55±0.14 | | 3.34±0.14 |
| 127 | 2005Sep01 | 53614.750 | 4.62±0.14 | | 3.54±0.15 |
| 128 | 2005Dec06 | 53711.420 | 4.63±0.14 | 19.14±1.61 | 4.09±0.17 |
| 129 | 2006Jun04 | 53891.490 | 3.71±0.11 | | 3.70±0.16 |
| 130 | 2006Sep01 | 53979.790 | 3.24±0.1 | | 3.49±0.15 |
| 131 | 2006Sep16 | 53994.780 | 3.12±0.09 | | 3.15±0.13 |
| 132 | 2006Sep17 | 53995.750 | 3.17±0.1 | | 3.40±0.14 |
| 133 | 2006Sep18 | 53996.700 | 3.20±0.1 | | 3.40±0.14 |
| 134 | 2006Sep19 | 53997.740 | 3.20±0.1 | | 3.44±0.14 |
| 135 | 2006Sep20 | 53998.750 | 3.18±0.1 | | 3.27±0.14 |
| 136 | 2006Oct28 | 54036.630 | 2.87±0.09 | 17.41±1.46 | 3.18±0.13 |
| 137 | 2006Oct29 | 54037.640 | 2.82±0.08 | | 3.40±0.14 |
| 138 | 2006Nov01 | 54040.650 | 2.84±0.09 | | 3.30±0.14 |
| 139 | 2006Dec18 | 54088.210 | 2.99±0.09 | | 3.31±0.14 |
| 140 | 2007Jan12 | 54112.640 | 3.46±0.1 | 12.16±1.02 | 3.18±0.13 |
| 141 | 2007Jan13 | 54113.650 | 3.32±0.1 | 14.72±1.24 | 3.14±0.13 |
| 142 | 2007Feb12 | 54143.610 | 3.92±0.12 | 15.86±1.33 | 3.35±0.14 |
| 143 | 2007May22 | 54242.890 | 4.89±0.15 | | 3.72±0.16 |
| 144 | 2007May23 | 54243.880 | 5.07±0.15 | | 3.97±0.17 |
| 145 | 2007Jun20 | 54271.870 | 4.85±0.15 | | 3.83±0.16 |
| 146 | 2007Jun21 | 54272.960 | 4.72±0.14 | | 4.20±0.18 |
| 147 | 2007Aug09 | 54321.820 | 3.73±0.11 | | 3.96±0.17 |
| 148 | 2007Aug10 | 54322.810 | | 12.46±1.05 | |
| 149 | 2007Aug11 | 54323.800 | 3.78±0.11 | | 4.06±0.17 |
| 150 | 2007Aug12 | 54324.810 | 3.81±0.11 | | 3.94±0.17 |
| 151 | 2007Sep03 | 54346.730 | 4.11±0.12 | 15.30±1.28 | 4.28±0.18 |
| 152 | 2007Sep04 | 54347.720 | 4.09±0.12 | | 4.30±0.18 |
| 153 | 2007Sep05 | 54348.720 | 4.10±0.12 | | 4.08±0.17 |
| 154 | 2007Oct16 | 54389.690 | 3.91±0.12 | 13.96±1.17 | 4.34±0.18 |
| 155 | 2007Oct18 | 54391.630 | 4.05±0.12 | | 4.42±0.19 |
| 156 | 2007Nov02 | 54406.650 | 3.80±0.11 | | 4.10±0.17 |
| 157 | 2007Nov04 | 54408.610 | 3.78±0.11 | 12.81±1.08 | 4.28±0.18 |
| 158 | 2007Nov08 | 54412.690 | 3.67±0.11 | | 4.12±0.17 |

**Notes.** – Col.(1): Number. Col.(2): UT-date. Col.(3): Julian date. Col.(4): $F$(cont), the continuum flux at 5100 Å (for Z=0) (in units of $10^{-15}$ erg s$^{-1}$ cm$^{-2}$ Å$^{-1}$), and $\varepsilon_{cont}$, the estimated continuum flux error. Col.(5): $F$(H$\alpha$), the H$\alpha$ total flux (in units of $10^{-13}$ erg s$^{-1}$ cm$^{-2}$) in regions (6740-7160)Å, and $\varepsilon_{H\alpha}$, the H$\alpha$ flux error. Col.(6): $F$(H$\beta$), the H$\beta$ total flux (in units of $10^{-13}$ erg s$^{-1}$ cm$^{-2}$) in regions (4909-5353)Å, and $\varepsilon_{H\beta}$, the H$\beta$ flux error.

**Table 3.** The estimated mean flux errors of the Hα and Hβ lines and their segments.

| Line | Region [Å] | σ±e [%] |
|------|-----------|---------|
| Hβ | 4909–5353 | 4.3±2.4 |
| Hα | 6740–7160 | 8.4±6.3 |
| *Line wings and core* | | |
| Hβ-blue wing | 4977–5097 | 4.2±3.0 |
| Hβ-core | 5097–5165 | 3.8±2.6 |
| Hβ-red wing | 5165–5268 | 4.8±3.0 |
| Hα-blue wing | 6720–6882 | 7.2±5.6 |
| Hα-core | 6882–6974 | 10.5±8.2 |
| Hα-red wing | 6974–7136 | 7.9±5.4 |
| *Blue and red part of line* | | |
| Hβ-blue part | 4977–5131 | 4.0±2.9 |
| Hβ-red part | 5131–5268 | 4.3±2.8 |
| Hα-blue part | 6720–6928 | 7.8±6.2 |
| Hα-red part | 6928–7136 | 8.8±6.4 |

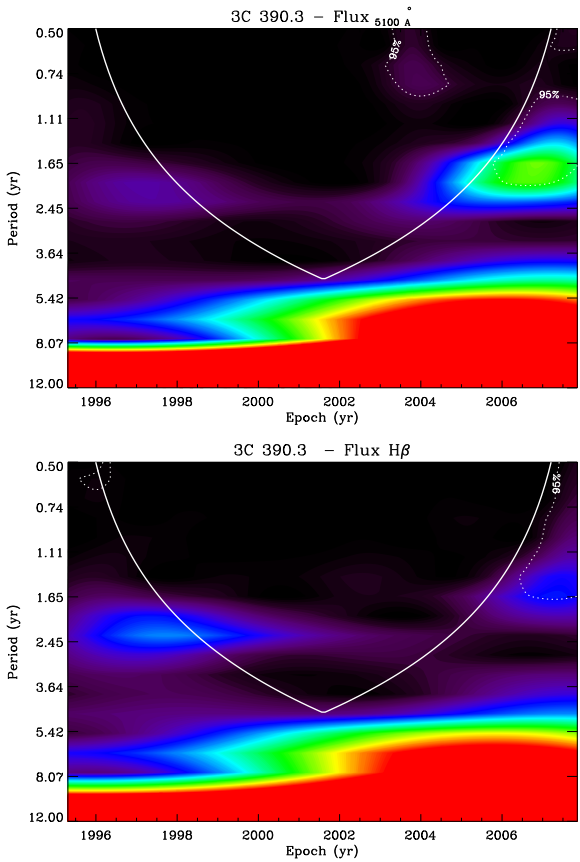

**Fig. 7.** The continuous wavelet power spectra of the long-term light curve of the optical continuum at 5100 Å(top) and in the Hβ (bottom) fluxes. The region below the thick dashed white line indicates the COI where edge effects resulting from the finite length of the time series, become important. Therefore, periods inside the COI might be dubious. The dashed contours indicate the 95 per cent confidence level. Redder colors corresponding to stronger signals.

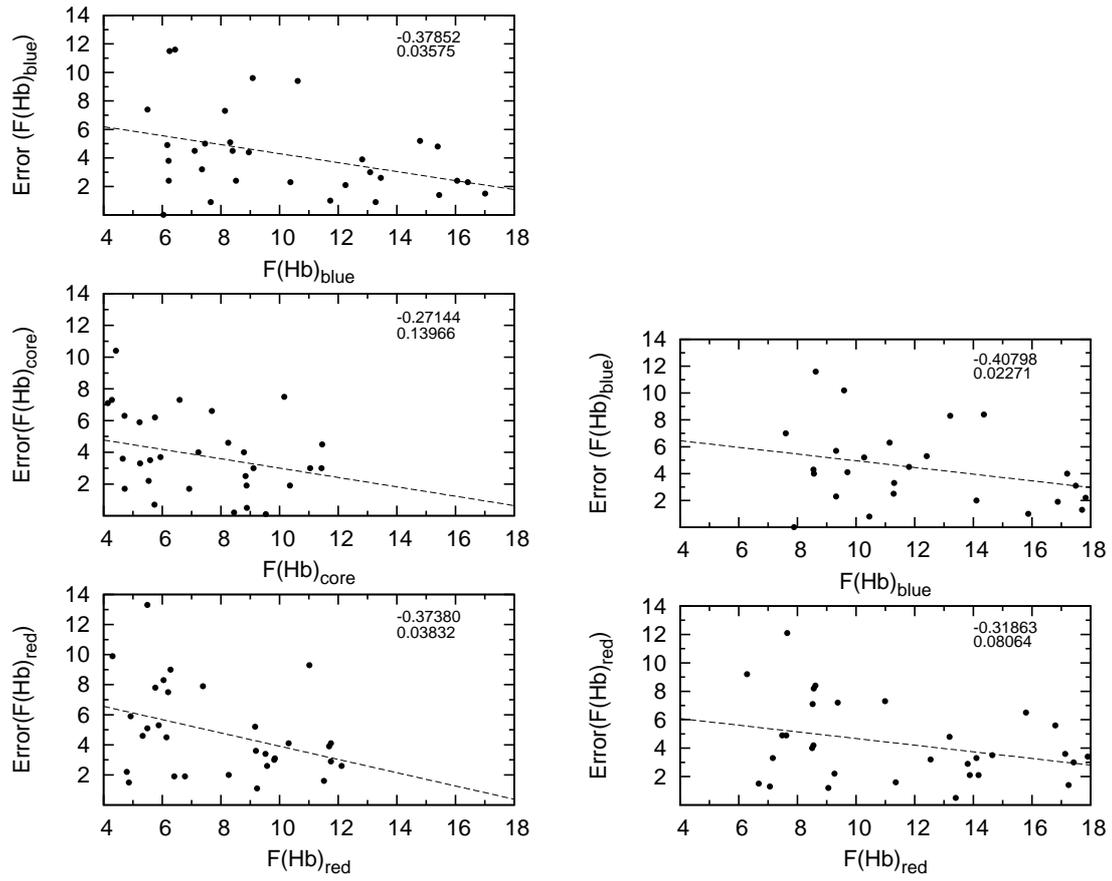

**Fig. 4.** The estimated errors in the flux measurements as functions of the corresponding flux, in case we divided lines in 3 parts (left panels) and 2 parts (right panels). The correlation coefficient and the p-value are given in the upper right corner.

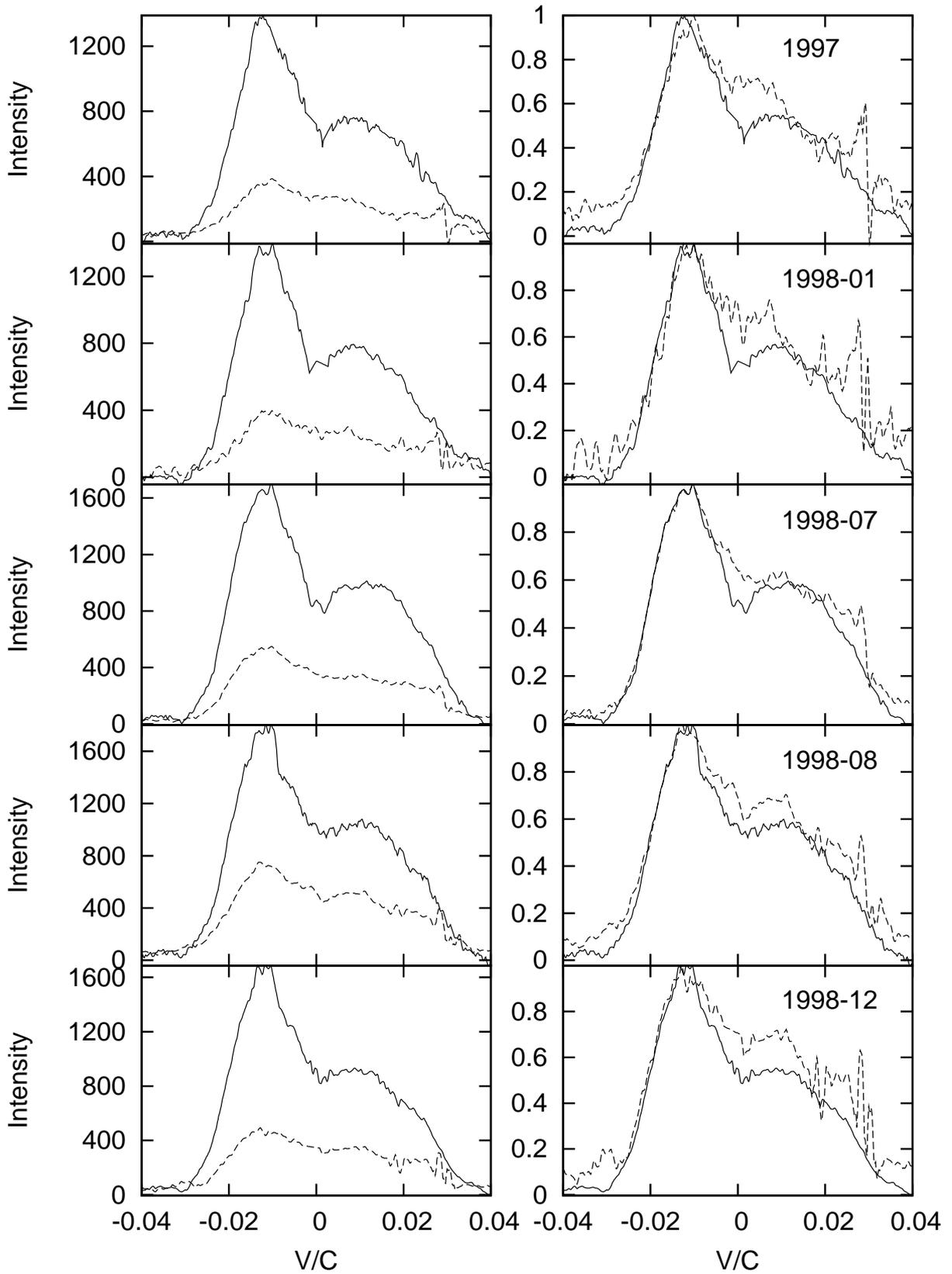

**Fig. 11.** The comparison of the Hα (solid line) and Hβ (dashed line) intensity (left) and their normalized profiles (right).

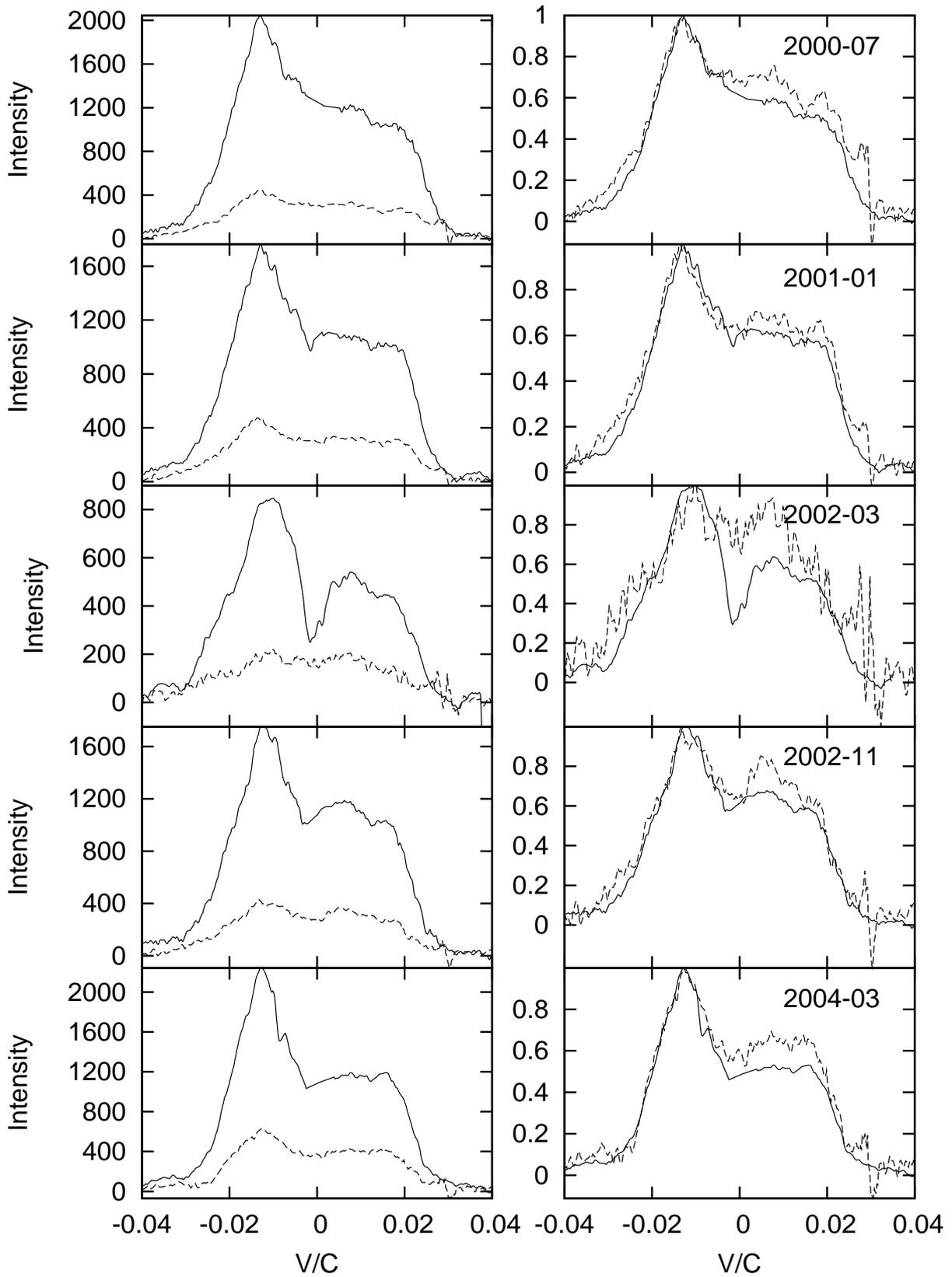

**Fig. 11.** Continued.

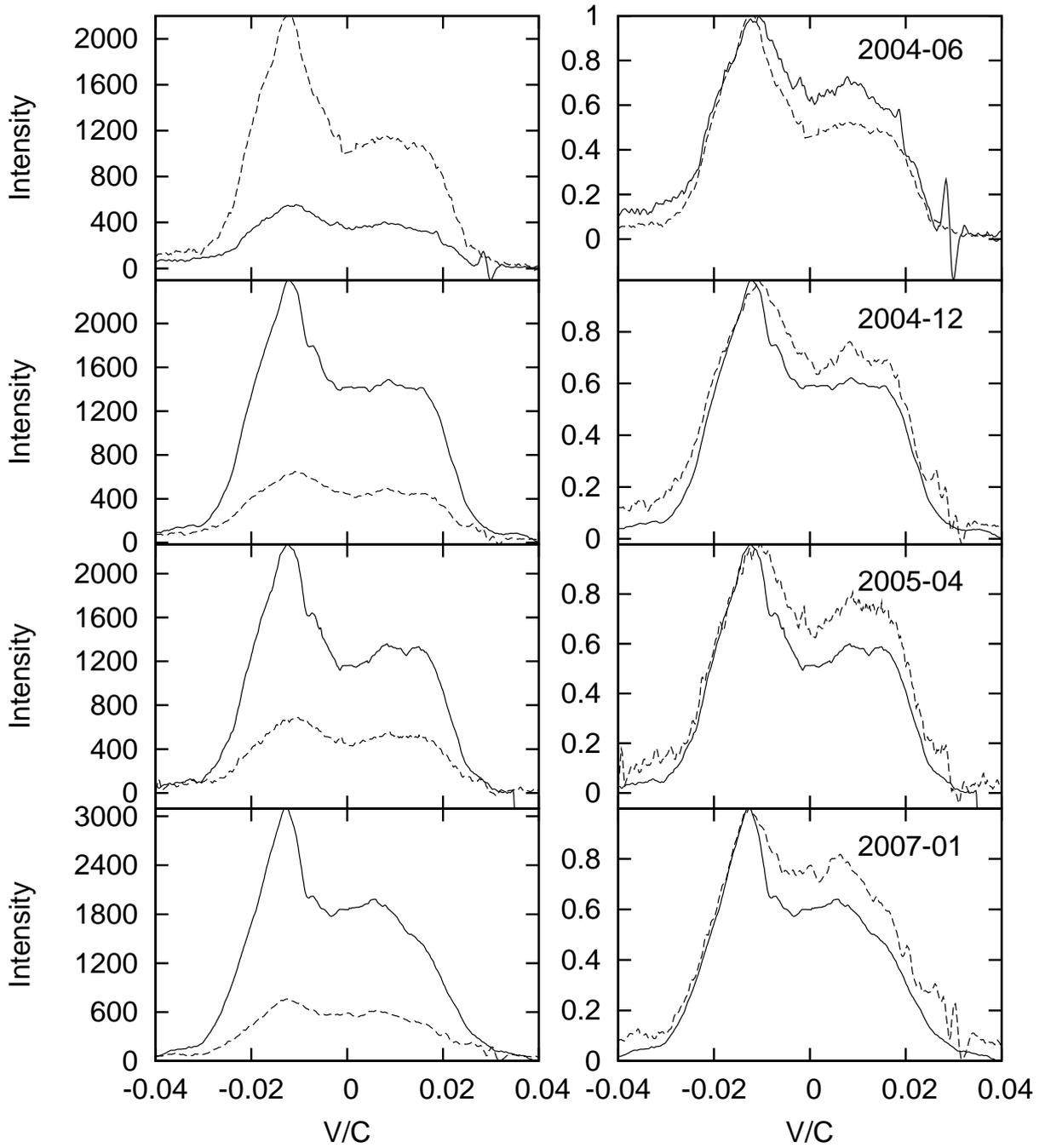

**Fig. 11.** Continued.

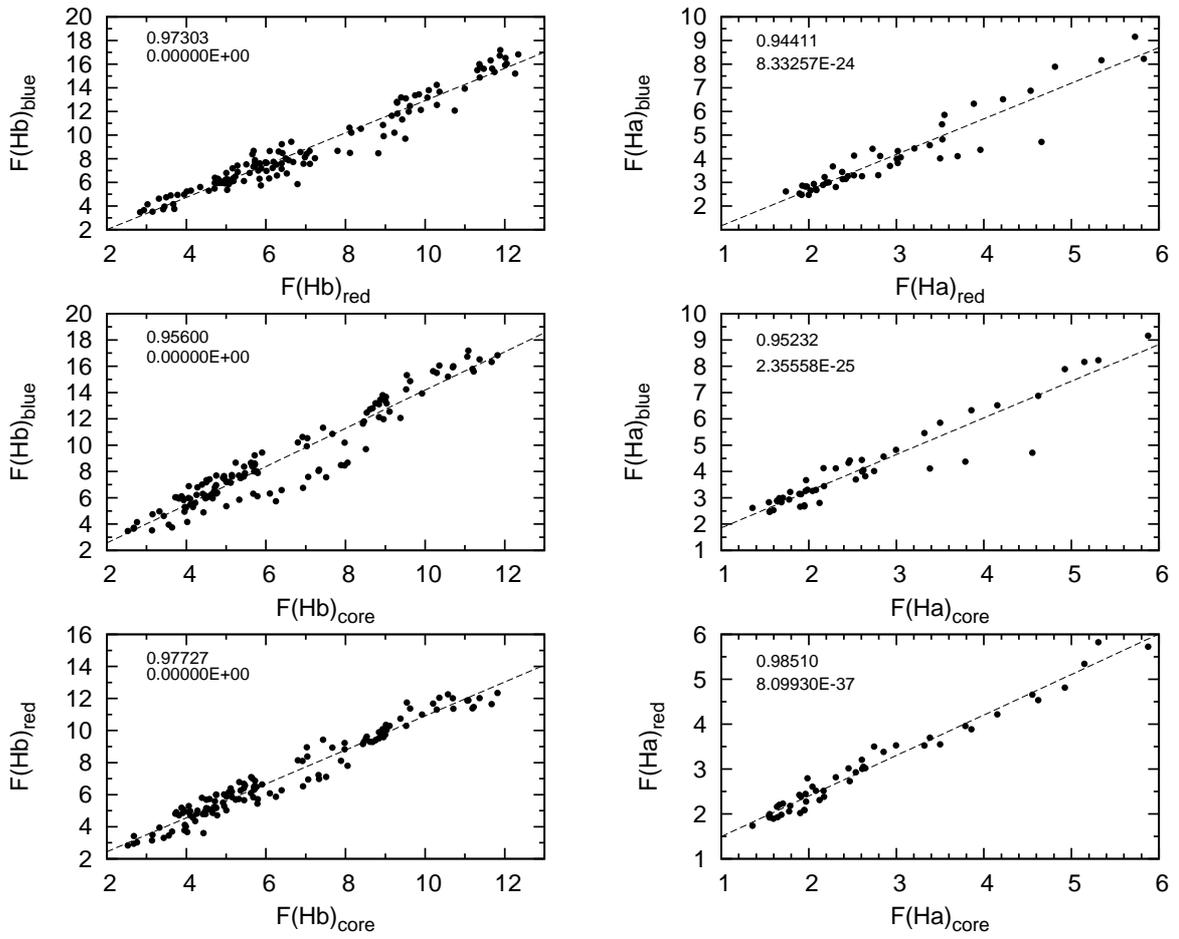

**Fig. 16.** Relationships between the H$\beta$ (left) and H$\alpha$ (right) line segments. The H$\beta$ flux is given in units $10^{-14}$erg m$^{-2}$ s$^{-1}$ and H$\alpha$ in $10^{-13}$erg m$^{-2}$ s$^{-1}$. The correlation coefficient and the p-value are given in the upper left corner.

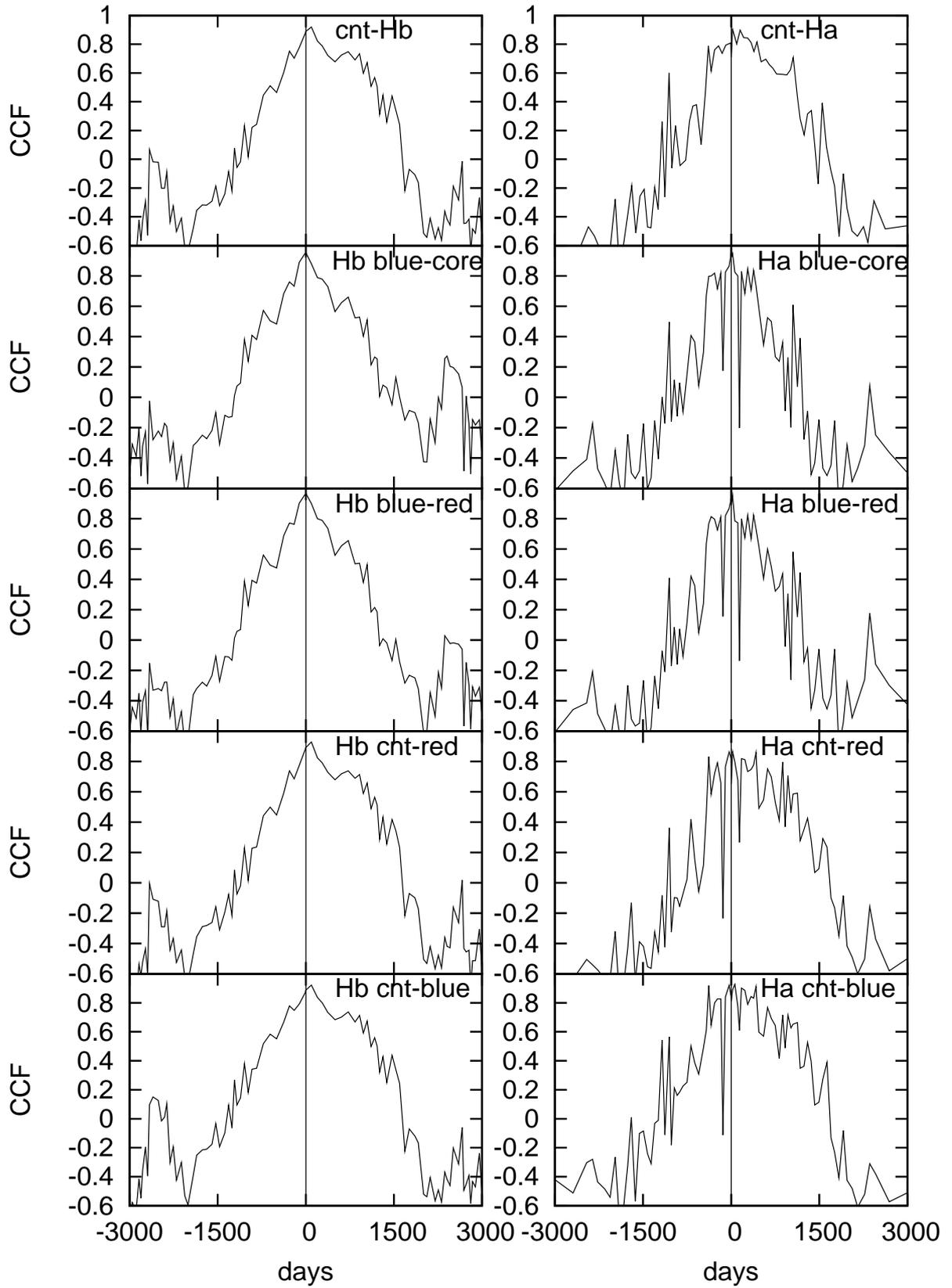

**Fig. 17.** CCFs for Hβ (left) and Hα (right).

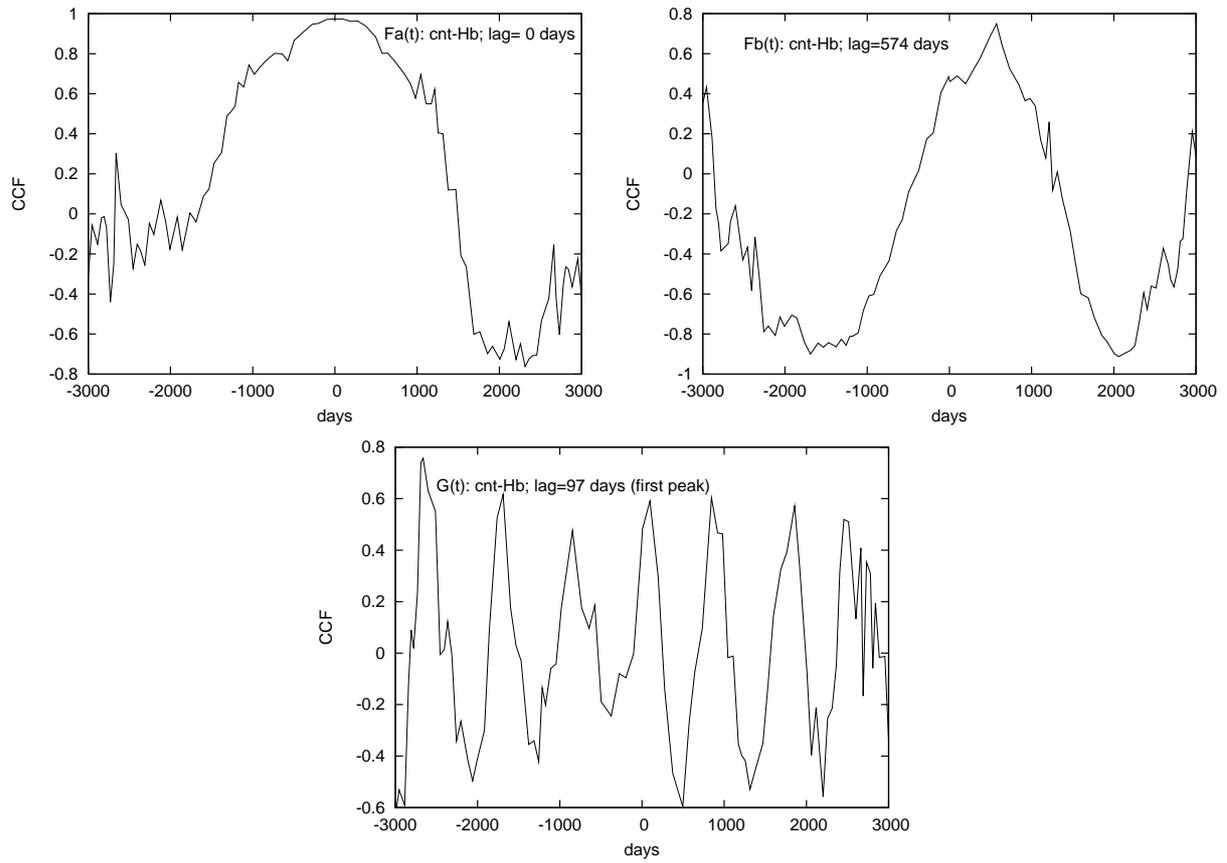

**Fig. 18.** CCFs between continuum and line functions $F_A$ (top), $F_B$ (middle) and G (down).

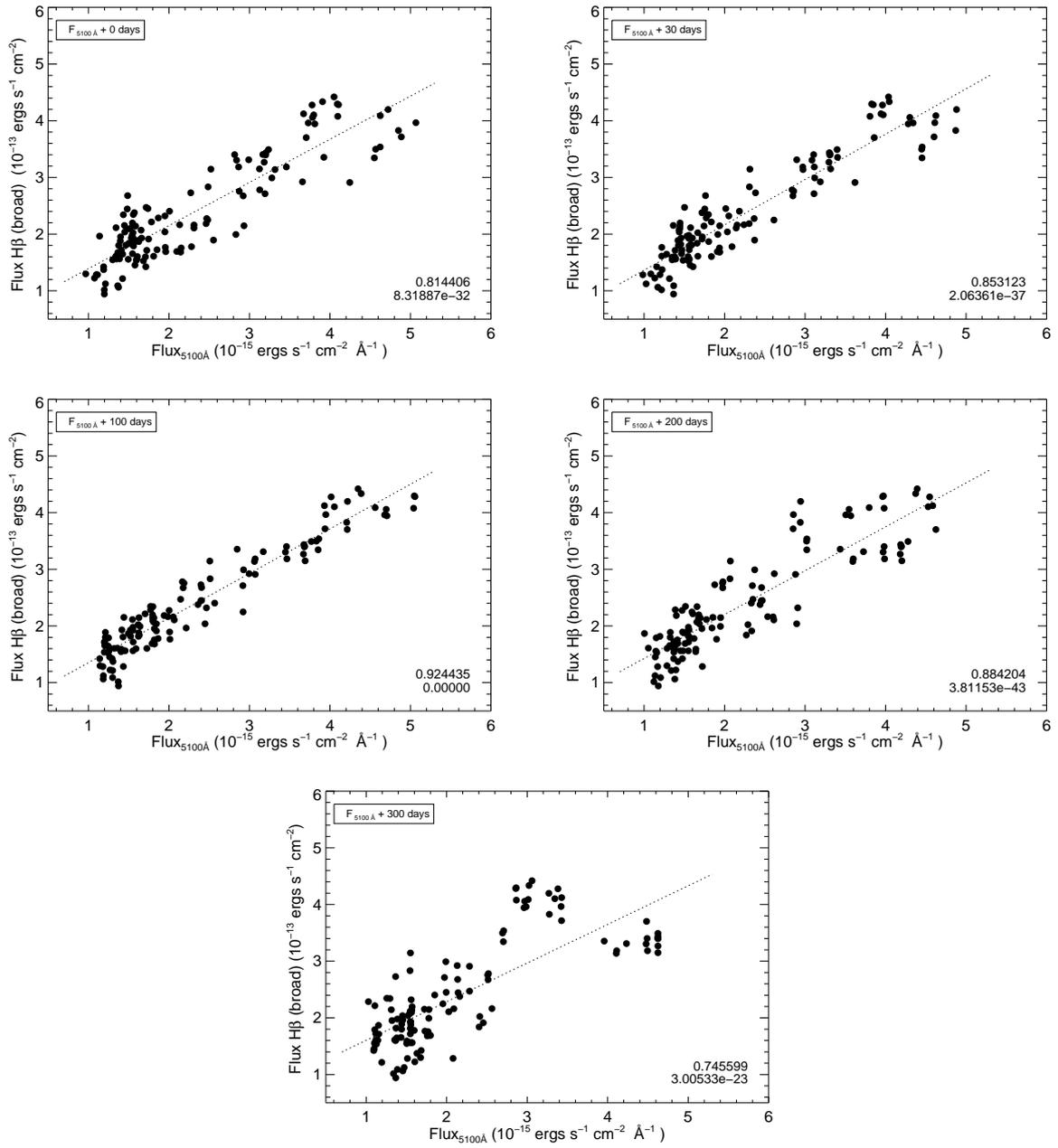

**Fig. 19.** Hβ vs. continuum correlations - with and without lags. Shifts for the continuum are given in the top left and the correlation coefficient and the p-value are in the bottom right.

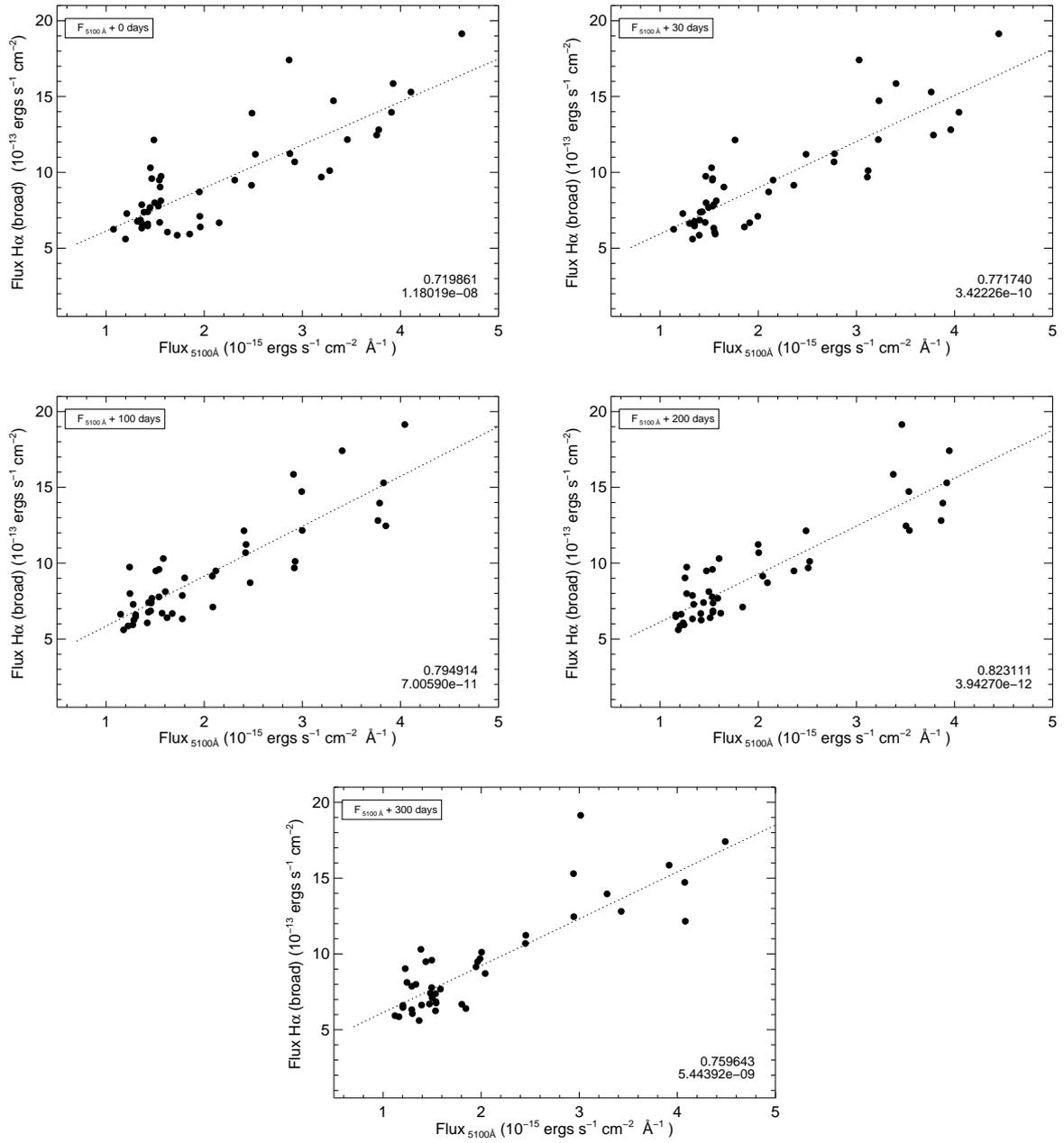

**Fig. 20.** Hα vs. continuum correlations - with and without lags. Shifts for the continuum are given in the top left and the correlation coefficient and the p-value are in the bottom right.